\documentclass[twoside]{article}

\usepackage[preprint]{aistats2026}
\usepackage{hyperref}
\usepackage{natbib}

\usepackage{amsthm}
\usepackage{bm}
\usepackage{amssymb}
\usepackage{booktabs}
\usepackage{microtype}
\usepackage{float}
\usepackage{graphicx}
\usepackage{subfigure}
\newtheorem{theorem}{Theorem}

\newtheorem{corollary}{Corollary}[theorem]  

\newcommand{\T}{\boldsymbol{T}}
\newcommand{\ub}{\boldsymbol{u}}
\newcommand{\Sb}{\boldsymbol{S}}
\newcommand{\y}{\boldsymbol{y}}
\newcommand{\Y}{\boldsymbol{Y}}
\newcommand{\R}{\boldsymbol{R}}
\newcommand{\1}{\boldsymbol{1}}

\DeclareMathOperator*{\argmin}{arg\,min}
\begin{document}

%

%

\twocolumn[

\aistatstitle{BASTION: A Bayesian Framework for Trend and Seasonality Decomposition}

\aistatsauthor{ Jason B. Cho \And David S. Matteson}

\aistatsaddress{
Department of Statistics and Data Science\\
Cornell University\\
Ithaca, United States
\And
Department of Statistics and Data Science\\
Cornell University\\
Ithaca, United States
} ]

\begin{abstract}
 We introduce BASTION (Bayesian Adaptive Seasonality and Trend DecompositION), a flexible Bayesian framework for decomposing time series into trend and multiple seasonality components. We cast the decomposition as a penalized nonparametric regression and establish formal conditions under which the trend and seasonal components are uniquely identifiable, an issue only treated informally in the existing literature. BASTION offers three key advantages over existing decomposition methods: (1) accurate estimation of trend and seasonality amidst abrupt changes, (2) enhanced robustness against outliers and time-varying volatility, and (3) robust uncertainty quantification. We evaluate BASTION against established methods, including TBATS, STR, and MSTL, using both simulated and real-world datasets. By effectively capturing complex dynamics while accounting for irregular components such as outliers and heteroskedasticity, BASTION delivers a more nuanced and interpretable decomposition. To support further research and practical applications, BASTION is available as an R package at \url{https://github.com/Jasoncho0914/BASTION}.
\end{abstract}

\section{Introduction}
\label{sec:intro}
Time-series decomposition is a powerful analytical method used to break down a univariate time-series into its constituent components, such as trend and seasonality. Decomposition reveals deeper insights into underlying long or medium-term trends that may not be immediately apparent. Historically, the most prominent applications of the method have been on economic reporting by government agencies, where it is used to adjust for seasonality when publishing key macroeconomic indicators such as the Consumer Price Index (CPI) \citep{App_cpi} and unemployment rates \citep{App_unemp}. 
Beyond economics, time-series decomposition has also been widely applied in climate science \citep{App_cli1,App_cli2}, epidemiology \citep{App_Epid1,App_Epid2}, and business and management \citep{app_biz3,app_biz1,app_biz2}, where seasonal effects play a crucial role in data analysis. 

More recently, time-series decomposition has received significant attention, as it has been shown to improve downstream task such as forecasting~\citep{rw_fcast3}. State-of-the-art time-series forecasting models, including TBATS \citep{tbats}, ETSformer \citep{Salesforce}, Prophet \citep{prophet}, DeepFS \citep{Amazon}, Autoformer \citep{autof}, TimeMixer \citep{rw_fcast2} and MSD-mixer \citep{rw_fcast1}, explicitly incorporate seasonal decomposition in their model architectures. Despite this widespread adoption, formal analysis of identifiability and principled uncertainty quantification for each component remain limited in the existing literature.

In this paper, we present Bayesian Adaptive Seasonality Trend decompositION (BASTION), a flexible Bayesian framework for trend-seasonality decomposition. A key innovation in BASTION is the use of a global–local shrinkage prior, which reduces noise while preserving salient signals, enabling locally adaptive yet smooth estimation of trend and seasonal components. The Bayesian formulation further enables the joint estimation of components that are typically treated outside the decomposition model. Such components include outliers or time-varying volatility. In addition, BASTION provides coherent uncertainty quantification for each component, an aspect that most existing methods lack. Beyond methodology, we develop the theoretical foundations required for reliable decomposition. Specifically, we establish conditions for identifiability of trend and seasonal components under penalized regression formulations.

The remainder of the paper is organized as follows. Section 2 reviews related work on time-series decomposition and global-local shrinkage priors. Section 3 formulates decomposition as a penalized regression problem and develops theoretical results on identifiability of decomposition components. Section 4 presents the Bayesian formulation of the model, including the use of global–local shrinkage priors and posterior inference. Section 5 evaluates the proposed method through extensive simulation studies and real data applications, including airline passenger counts and electricity demand. 

\section{Related Work}
Much of the time-series decomposition literature treats decomposition primarily as a preprocessing or exploratory tool, rather than as a statistical inference problem in its own right. As a result, existing methods often exhibit several limitations, including limited ability to accommodate multiple seasonal patterns, difficulty adapting to abrupt changes in the trend, lack of principled uncertainty quantification, inability to handle heteroskedastic noise, and limited robustness to outliers. While individual approaches address subsets of these challenges, no existing method resolves all of them simultaneously. 

Foundational work on seasonality and trend decomposition includes Seasonal–Trend decomposition using LOESS (STL)\citep{STL} and structural time-series modeling approach by \citet{sts}. Numerous extensions and alternatives have since been proposed. For example, Multiple STL (MSTL)\citep{MSTL} extends STL to accommodate multiple seasonalities, but does not provide uncertainty quantification and remains sensitive to sudden shifts in the trend. Seasonality–Trend decomposition with Regression (STR)\citep{STR} supports complex seasonal patterns and incorporates uncertainty quantification, yet remains limited in its ability to capture abrupt changes. Robust STL~\citep{robostSTL} improves robustness to outliers and accommodates multiple seasonal patterns and abrupt changes in the trend, but does not provide uncertainty quantification. Moreover, none of the existing approaches explicitly model heteroskedastic noise, which is commonly observed in time-series data across various disciplines such as finance~\citep{CV2, SVop, SVop2, SVop3}, epidemiology~\citep{Epidem, Epidem2}, geophysics~\citep{geophysics2, geophysics1}, and environmental science~\citep{climate1, climate2, climate3}.

We address these challenges by adopting a Bayesian framework and incorporating global–local shrinkage priors. Global–local shrinkage priors have emerged as a powerful class of continuous shrinkage priors for high-dimensional regression~\citep{horseshoe, horseshoe_P} and have recently been extended to time-series settings, including Bayesian smoothing~\citep{dsp, dspCount}, time-series regression~\citep{tgamma}, change-point detection~\citep{abco, horseshoe_cp}, and volatility estimation~\citep{ASV, dspinVolatility}. To the best of our knowledge, however, these priors have not previously been applied to time-series decomposition; the present work fills this gap by incorporating global–local shrinkage priors within the framework.

\section{Decomposition as a Penalized Regression Problem}\label{sec:decomp}
We first frame the decomposition problem in the language of penalized regression. This provides a familiar framework for situating our work within the broader nonparametric regression literature and highlights connections with existing approaches by \cite{sts} and \cite{STR}. Given a length $N$ time-series $\y:=\{y_{t}\}_{t=1}^{N}$, an additive observation equation structure is assumed, consisting of the trend component $\T := \{T_{t}\}_{t=1}^{N}$, $P$ seasonal components $\Sb_i := \{S_{i,t}\}_{t=1}^{N}$ with corresponding seasonal periods $\{k_i\}_{i=1}^{P}$, and the gaussian remainder $\R := \{R_{t}\}_{t=1}^{N}$,
\begin{align}
	&y_{t} =  T_{t} + \sum_{i=1}^{P}S_{i,t}  + R_{t}, & &[R_{t}|\sigma^2_y] \sim N(0,\sigma^2_{y}).
	\label{obs}
\end{align}
This is a high-dimensional problem: $N(P+1)$ unknown parameters must be estimated from only $N$ observations. Without additional structure, the maximum likelihood estimate is not identifiable. Specifically, structure on each latent component must be imposed.

Define $D_{T}$ and $D_{S,i}$ be the linear operators for trend and the $i$-th seasonal component, and  $\lambda_{T},\lambda_{S,i} > 0$ be the hyperaparameter controlling the penalty. The norm of the linear operation of each component is further penalized. With this additional regularization, the maximum likelihood estimates are written as:
\begin{equation}
    \begin{aligned}\label{obj1}
	\hat{\T},\hat{\Sb_1},&\ldots,\hat{\Sb_P}  = \argmin_{\T,\Sb_1,\ldots,\Sb_P}\bigg\{\|Y - \T - \sum_{i=1}^{P}\Sb_i\|^2_2 \\
	 & + \lambda_{T} \|D_{T}\T\|^2_2 +\sum_{i=1}^{P}\lambda_{S,i}\|D_{S,i}\Sb_i\|^2_2 \bigg \}.
    \end{aligned}
\end{equation}
A natural choice of the linear operator, $D_T$, for the trend is the second-difference operator, often denoted by $\Delta^2$, where $\Delta^2 T_t  = T_{t+1} - 2T_{t} + T_{t-1}$. The use of second differencing for smoothing is well established in the literature, with applications in smoothing splines \citep{Sspline}, Hodrick-Prescott filter \citep{HP}, $\ell_1$-trend filter \citep{l1tf1,l1tf2}, and Bayesian Trend filters \citep{btf,dsp}.

For the seasonality, \cite{sts} proposes the seasonal recurrence operator: $S_{t} = -\sum_{m=1}^{k-1}S_{t-m}$, where $k$ is the length of the seasonality or $(I + B + B^2 + \ldots + B^{k-1})$ as a matrix operation with $B$ denoting the backshift operator. Another natural choice would be the $k$-th order backshift operator $(1-B^{k})$, where $k$ is the length of the cycle, thereby penalizing, $S_{i,t+k} - S_{i,t}$. While not used for time-series decomposition, such operation shows up in a classical time-series modeling literature such as SARIMA type models. 

\subsection{Identifiability}
When each component is directly parameterized and estimated, different allocations of variation across components can yield the same fit to the data, leading to identifiability issues. Existing work such as \cite{STR} has addressed this only informally by imposing centering constraints on seasonal terms. To our knowledge, this is the first work to formalize identifiability conditions in the context of time-series decomposition. In particular, we show that even after penalizing each component with the operators discussed above, the solution to Equation~\ref{obj1} need not be uniquely identified. 

The first order condition of the Equation~\ref{obj1} gives us the following set of equations:
\begin{align*}
		&A_{0}\hat{\T}  =  \Y - \sum_{i=1}^{P}\hat{\Sb_i}\\
		&A_{1}\hat{\Sb_1}  =  Y -\hat{\T} - \sum_{j \neq 1}\hat{\Sb_{j}}\\
		& \ldots \\
		&A_{P}\hat{\Sb_P}  =  Y -\hat{\T} - \sum_{j \neq P}\hat{\Sb_{j}},
\end{align*}
where $A_{0} = (I + \lambda_{0} L_{0})$, $L_{0} = (D'_{\T} D_{\T})$, and $A_{j} = (I + \lambda_{j} L_{j})$, $L_{j} = (D_{S,j}'D_{S,j})$ for $j  = 1,\ldots P$. This means that the identifiability of the solution hinges on whether the block matrix $H$,
\[H = \begin{bmatrix}
		A_{0} & I & \ldots & I \\
		I     & A_{1} & \ldots & I\\
		\vdots & \vdots & \ddots &\vdots \\
		I & I  & \ldots & A_{P}
	\end{bmatrix},
\] arising from this first-order conditions is full-rank, thus invertible.
\begin{theorem}[Identifiability Condition]
The solution in Equation~\ref{obj1} is uniquely identifiable if and only if 
\begin{equation*}
\begin{aligned}
   &\dim\ker(D_{T}) + \sum_{j=1}^P \dim\ker(D_{S,j}) \\
   &= \dim\!\Big(\ker(D_{T}) + \sum_{j=1}^P \ker(D_{S,j}) \Big)
\end{aligned}
\end{equation*}
Equivalently, $\ker(D_T),\ker(D_{S,1}),\dots,\ker(D_{S,P})$ are linearly independent subspaces.
\begin{proof}
	The argument relies on expanding the first-order conditions and analyzing the induced sum map on the kernels of the $D_T$ and $D_j$. A full derivation is provided in Appendix~\ref{pf:id}.
\end{proof}
\end{theorem}

This theorem shows that the only possible non-identifiability comes from directions simultaneously unpenalized by each operators. The second differencing for the trend and recurrent seasonal operator as suggested by \cite{sts} or seasonal differencing operator for the seasonality results in a unique solution.
\begin{corollary}[Single seasonality]\label{c11}
For $P=1$, let $D_T$ be the second–difference operator and $D_{S,1}$ be either the seasonal recurrent operator $(1+B+\cdots+B^{k-1})$, or the seasonal differencing operator $(1-B^k)$.Then the decomposition is uniquely identifiable under the seasonal recurrent operator, and identifiable only up to an additive constant under the seasonal differencing operator.
\begin{proof}
Since $\dim \ker(D_T)=2$ and $\dim \ker(D_{S,1})=k$, we compute their intersections. For the recurrent operator, $\ker(D_T)\cap\ker(D_{S,1})={\mathbf{0}}$, so the decomposition is unique. For the differencing operator, $\ker(D_T)\cap\ker(D_{S,1})=\operatorname{span}(\mathbf{1})$, so the decomposition is identifiable only up to a constant shift. Full details are given in Appendix~\ref{pf:c11}.
\end{proof}
\end{corollary}
However, identifiability is lost once multiple seasonal components are introduced. The corollary below shows that the identifiability of a model depends on the greatest common denominator of the seasonal lengths when either seasonal recurrent or differencing operator is applied to penalize each seasonal component.
\begin{corollary}[Multiple seasonalities]\label{cp13}
For $P>1$, with $D_T$ given by the second differencing and each $D_{S,j}$ given by either 1) seasonal differencing or 2) seasonal recurrent operator, the decomposition is not uniquely identifiable.
\begin{proof}
We show that the nullity of the design matrix,
\begin{align*}
   \text{nullity}&(H_{R}) = \sum_{i<j}\gcd(k_i,k_j) \\
   - &\sum_{i<j<l}\gcd(k_i,k_j,k_l) + \dots - \gcd(k_1,\ldots,k_P), 
\end{align*}
   for the seasonal recurrence operator and 
$$ \text{nullity}(H_S) = 1+\text{nullity}(H_{R}),$$
for the seasonal differencing operator. $k_{1},\ldots,k_{P}$ are the length of each season and $\gcd$ is the greatest common denominator (Appendix~\ref{pf:c12}).
\end{proof}
\end{corollary}
\subsection{Multiple Regularization Operators}
The identifiability issues arising with multiple seasonalities stem from large null spaces left unpenalized by any single operator. A natural way to address this problem is to penalize each component with multiple linear operators simultaneously, thereby reducing the intersection of their null spaces. For instance, one may penalize a seasonal component both by second differencing (to enforce smoothness) and by seasonal differencing (to enforce periodicity). Under this framework, the estimator becomes
\begin{equation}
    \begin{aligned}
		\hat{\T},\hat{\Sb_1},&\ldots,\hat{\Sb_P}  = \argmin_{\T,\Sb_1,\ldots,\Sb_P}\bigg\{\|Y - \T - \sum_{i=1}^{P}\Sb_i\|^2_2 \\
		&+ \sum_{m=1}^{q_T}\lambda^m_{T} \|D^m_{T}\T\|^2_2 \\
		&+ \sum_{m=1}^{q_S}\lambda^{m}_{S,i}\sum_{i=1}^{P}\|D_{S,i}^m\Sb_i\|^2_2 \bigg \}.
    \end{aligned}\label{obj2}
\end{equation}
$q_T$ and $q_S$ represent the number of penalties for the trend and seasonal component respectively. The hyperparameters, $\lambda_T^m,\lambda_{S,i}^m > 0$ are strictly positive scalar controlling the degree of penalty applied to each operator. 
\begin{theorem}[Identifiability with Multiple Penalties]\label{t2}
Let $V_0=\bigcap_{m=1}^{q_T}\ker(D_T^m)$ and 
$V_j=\bigcap_{m=1}^{q_S}\ker(D_{S,j}^m)$ for $j=1,\dots,P$. 
Then, the solution in Equation~\ref{obj2} is identifiable if and only if
\[
\sum_{j=0}^P \dim V_j=\dim\Big(\sum_{j=0}^P V_j\Big),
\]
Equivalently, the subspaces $V_0,\dots,V_P$ are linearly independent subspaces.
\begin{proof}
	Similar proof is performed as in Theorem 1. A full derivation is provided in Appendix~\ref{pf:idmult}.
\end{proof}
\end{theorem}
\begin{corollary}[Mixed differencing 1]\label{c21}
For $P \ge 1$, consider the trend penalized by the second-difference operator $D_{T}$ and each seasonal component $\Sb_{j}$ penalized by either
\begin{enumerate}
    \item a second-difference operator $D_{S,j}^1$ and a seasonal differencing operator $D_{S,j}^2$ with an additional constraint that $\1^T\Sb_{j} = 0$ or 
    \item a second-difference operator $D_{S,j}^1$ and a seasonal recurrence operator $D_{S,j}^2$,
\end{enumerate}
the decomposition is uniquely identifiable.
\begin{proof}
By imposing both the second and seasonal differencing penalties, each seasonal component has only a constant, $\text{span}(\1)$, in its null space. The additional centering constraint $\mathbf 1^\top \Sb_j=0$ removes this from the null space, which leaves only the trend null space. The null space reduces to 0 when the seasonal recurrence and second-difference operator are imposed. Detailed derivation is in Appendix~\ref{pf:c21}.
\end{proof}
\end{corollary}
This result shows that in the multiple-seasonality setting, the identifiability issues arising under seasonal differencing and recurrence penalties can be resolved by additionally imposing a second-difference operator on each seasonal component. More generally, Theorem \ref{t2} provides a simple diagnostic: for any proposed set of operators, one can compute the associated null spaces $V_0,\ldots,V_P$ and check whether they are linearly independent.

\section{Bayesian Perspective}\label{bayP}
Building on this foundation, we now propose \textbf{BASTION} (Bayesian Adaptive Seasonal Trend decompositION), a unified Bayesian model for trend-seasonality decomposition. We extend the penalized least squares formulation described in Section~\ref{sec:decomp} by casting it in a Bayesian framework. Recall that $D_{T}$ and $D_{S,i}$ for $i  = 1,\ldots,P$ are the operators for the trend and seasonal components, respectively. We assign Gaussian priors:
\begin{align*}
	&[D_{T}\T |\sigma_{T},\sigma_y ] \sim N(0,\sigma^2_y\sigma^2_TI), \\ 
	&[D_{S,i}\Sb_i|\sigma_{s,i},\sigma_y] \sim N(0,\sigma^2_{y}\sigma^2_{S,i}I),
\end{align*}
together with the observation equation described in Equation~\ref{obs}. Under this specification, the posterior mode of $\T,\Sb_i,\ldots,\Sb_P$ is identical to the penalized estimator described in Equation~\ref{obj1}, with penalty weights $\lambda_T = 1/\sigma^2_T$ and $\lambda_{S,i} =1/\sigma^2_{S,i}$. 

The penalized regression with multiple operators as described in Equation~\ref{obj2} also admits a Bayesian interpretation. In this setting, the appropriate prior is not a product of Gaussians but a Gaussian random field, since the quadratic forms are not well defined as independent priors. Specifically:
\begin{align*}
	&P(\T)  \propto \exp\bigg\{-\frac{1}{2\sigma^2_y}\T'\bigg(\sum_{m=1}^{q_T} \frac{1}{\sigma^2_{T,m}}(D_{T}^m)'D_{T}^m \bigg)\T \bigg\}\\
	&P(\Sb)  \propto \exp\bigg\{-\frac{1}{2\sigma^2_y}\sum_{i=1}^{P}\Sb_i'\bigg(\sum_{m=1}^{q_T} \frac{1}{\sigma^2_{S,i,m}}(D_{S,i}^m)'D_{S,i}^m \bigg)\Sb_i \bigg\}.
\end{align*}
The posterior mode under this specification also reduces to the penalized estimator described in Equation~\ref{obj2} with $\lambda_T^m = 1/\sigma^2_{T,m}$ $\lambda_{S,i}^m = 1 / \sigma^2_{S,i,m}$.

The Bayesian formulation not only recovers the penalized estimators from the previous sections, but also broadens the framework in several directions. It provides coherent uncertainty quantification through posterior credible intervals, accommodates various prior structures that can encode smoothness, sparsity, or other domain knowledge and also incorporates additional components such as heterogeneous error variances \citep{kimetal} or outlier component \citep{abco}. 
\begin{table*}[!ht]
\centering
\resizebox{0.8\textwidth}{!}{%
\begin{tabular}{@{}llllll@{}}
\toprule
\multicolumn{1}{c}{}              & \multicolumn{1}{c}{\textbf{Trend}}       & \multicolumn{1}{c}{\textbf{Seasonality}}                                      & \multicolumn{1}{c}{\textbf{\begin{tabular}[c]{@{}c@{}}Seasonal\\ Length\end{tabular}}} & \multicolumn{1}{c}{\textbf{Remainder}}         & \multicolumn{1}{c}{\textbf{Outlier}} \\ \midrule
\textbf{DGP 1}             & Piecewise Linear                         & Fourier Pairs                                                                 & 12, 40                                                                                 & Constant                                   & FALSE                                \\
\textbf{DGP 2}             & Linear                                   & Piecewise Constant                                                            & 40                                                                                     & Constant                                   & FALSE                                \\
\textbf{DGP 3}             & Quadratic                                & Fourier Pair                                                                  & 50                                                                                     & Stochastic Volatility                      & FALSE                                \\
\textbf{DGP 4}             & Piecewise Linear                         & \begin{tabular}[c]{@{}l@{}}Piecewise Constant \\ \& Fourier Pair\end{tabular} & 12, 40                                                                                 & Stochastic Volatility                      & TRUE                                 \\
\toprule
\end{tabular}%
}
\caption{Descriptions of data generating processes (DGP) with regards to their trend, seasonality, seasonal lengths, remainder, and outliers. The sample size of each time-series is 500, and each DGP is replicated 1000 times with random coefficients for both the trend and seasonality components. Exact specification of each simulation schemes can be found in Appendic~\ref{App:simscheme} and example paths from each DGP can be found in Appendix~\ref{App:addfigs}.}
\label{tab:simulation_descriptions}
\end{table*}
\subsection{Global-Local Shrinkage Prior}\label{sec:gl}
As shown in Equations~\ref{obj1} and \ref{obj2}, existing frameworks impose a uniform penalty term $\lambda_\bullet$ on each operator, which can oversmooth sharp transitions or underpenalize noise. Global–local shrinkage priors address this limitation by coupling an overall shrinkage parameter that enforces stability with time-varying local parameters that selectively relax shrinkage when large changes occur. The defining feature of BASTION is the use of such global–local shrinkage priors for both trend and seasonal components, enabling estimates that evolve smoothly while still capturing abrupt shifts. Formally, we specify:
\begin{align*}
	&[D_{T}\T |\sigma_y,\eta_{T},\tau_T] \sim N(0,\sigma^2_y\tau^2_T\eta_{T}^2I), & & \\ 
	&[\tau_T] \sim p(\tau_T), \quad [\eta_{T}] \sim p(\eta_{T}),\\ 
	&[D_{S,i}\Sb_i|\sigma_y,\tau_{S,i},\eta_{S,i}] \sim N(0,\sigma^2_{y}\tau_{S,i}^2\mathbf{\eta}_{S,i}^2I),\\
	&[\tau_{S,i}] \sim p(\tau_{S,i}), \quad [\eta_{S,i}] \sim p(\eta_{S,i}),
\end{align*}
where each component is regularized by a global scalar parameter $\tau_{\bullet}$, controlling the overall shrinkage, paired with a time-varying local parameter \(\eta_{\bullet} := [\eta_{\bullet,1},\ldots\eta_{\bullet,N}]'\) that captures local variation. 

Several global-local shrinkage priors have been proposed, including the horseshoe prior \cite{horseshoe}, horseshoe+ \cite{horseshoe_P}, dynamic shrinkage processes \cite{dsp}, the regularized horseshoe \cite{regularized_hs}, and the triple Gamma prior \cite{tgamma}, each differing in shrinkage strength. Considering the large number of parameters to be estimated for time-series decomposition task, the horseshoe prior is chosen, as it requires the least number of parameters. Under this specification:
\begin{align*}
	&[\tau_{\bullet}] \sim C^{+}(0,1/N), & [\eta_{\bullet, t}] \stackrel{iid}{\sim} C^{+}(0,1).  
\end{align*}

\subsection{Explicit Outlier Detection}
Empirical time series frequently exhibit atypical observations due to measurement error or rare events. Existing methods typically handle outliers by pre-adjusting the observed series or by applying smoothing methods that are robust to outliers, rather than directly addressing them within the decomposition framework \citep{x12Arima,STR}. BASTION is the first decomposition method to explicitly model outliers through the use of an extreme shrinkage-inducing prior inspired by the approach in \cite{abco}. To capture large deviations at isolated time points, BASTION applies the horseshoe+ prior by \cite{horseshoe_P} to \(\zeta_t\).  By nesting two levels of half-Cauchy distributions, the horseshoe+ prior provides more aggressive shrinkage, allowing larger deviations from zero at a fewer number of locations when compared to the regular horseshoe prior. Specifically,
\begin{align*}
    &[\zeta_{t}|\sigma_y,\eta_{\zeta,t}] \sim N(0, \sigma_y^2 \eta^2_{\zeta,t}), \\
    & [\eta_{\zeta,t}|\tau_{\zeta},\xi_{\zeta,t}] \sim C^+(0, \tau_{\zeta} \xi_{\zeta,t}), \\
    &[\tau_{\zeta}] \sim C^+(0,1), \\
    & [\xi_{\zeta,t}] \stackrel{iid}{\sim} C^+(0,1).
\end{align*}
\subsection{Handling Heteroskedasticity}
While heteroskedastic noise, characterized by changing volatility over time, is often observed in data sets across various academic disciplines, no existing decomposition methods explicitly account for it. Another novel feature of BASTION is its ability to directly handle heteroskedasticity. To address this gap, BASTION introduces a stochastic volatility (SV) model to the remainder term. In addition to the overall variance parameter, $\sigma_y$, we introduce a time-varying variance component $\{\nu_{t}\}_{t=1}^{N}$, which follows a first order stochastic volatility model by \cite{kimetal}, defined as follows:
\begin{align*}
	&[R_{t}|\sigma_{y},\nu_{t}]\sim N(0,\sigma_y^2 \nu_t^2) \\
	&log(\nu^2_t) = \mu + \phi(log(\nu^2_{t-1}) - \mu) + \sigma_{\nu}\epsilon_{t} \\
	&[\epsilon_{t}] \stackrel{iid}{\sim} N(0,1). 
\end{align*}
Here, $\mu$ represents the mean log-variance, $\phi$ is a persistence parameter that controls how strongly current volatility depends on past values, and $\sigma_{\nu}$ dictates the variability of the log-variance process. By incorporating a time-varying volatility model, BASTION provides a more flexible decomposition that adapts to heteroskedastic structures in the data. This extension allows for more accurate trend and seasonality estimation, particularly in data with substantial fluctuations in residual variance.

\subsection{Implementation}
Fully Bayesian inference provides calibrated uncertainty quantification by sampling from the posterior distribution. General-purpose samplers such as the No-U-Turn Sampler (NUTS) \citep{nuts}, implemented in Stan \citep{stan}, are popular for complex models because of their automatic tuning and robust convergence properties \citep{hmcgood,hmcbad3}. However, for high-dimensional and structured time-series models like BASTION, these methods are often computationally infeasible due to poor scaling and slow mixing \citep{hmcbad1,hmcbad2}.

Instead, we adopt a Gibbs sampling strategy that scales linearly with the length of the time series, making it practical for large datasets if the goal is to sample from the exact posterior distribution. With the exception of the persistence parameter $\phi$, the persistence parameter for the SV model in the remainder term, all full conditional distributions are available in closed form and admit conjugate updates. Because all the state variables, $\{T_{t}\}$, $\{S_{t}\}$, and $\{\zeta_{t}\}$  are conditionally Gaussian and sampled using the multivariate Gaussian sampler of \citet{rue}, which is $\mathcal{O}(N)$. Parameters associated with the horseshoe prior are updated using the parameter-expanded Gibbs scheme of \citet{easyHSsampler}, while the stochastic volatility parameters in the remainder term are sampled with the All Without a Loop (AWOL) algorithm of \citet{Kastner_2014}. Importantly, both procedures, like the Gaussian state sampler, scale linearly with the number of observations. Detailed specification of the prior distribution as well as the conditional posterior distribution for Gibbs sampling is detailed in Appendix~\ref{App:gibbs}. Implementation of BASTION in \texttt{R} is available at \url{https://github.com/Jasoncho0914/BASTION}.

\section{Simulation Study}\label{sec:simulationstudy}
\subsection{Set Up}
In this section, BASTION is compared against existing multiple seasonalities decomposition models: MSTL by \cite{MSTL}, STR by \cite{STR}, and TBATS by \cite{tbats} across various simulation scenarios described in Table~\ref{tab:simulation_descriptions}. Exact simulation schemes are detailed in Appendix~\ref{App:simscheme}. DGP 1 through 4 exhibit characteristics such as trend discontinuities, seasonal discontinuities, heteroskedastic noise, or a combination of these with additive outliers. Note that DGP 3 and DGP 4 include heteroskedastic noise, with additive outliers additionally introduced in DGP 4. The comparisons are made in terms of their ability to accurately extract trend, seasonality, and signal (trend and seasonality combined), by measuring the mean squared error (MSE). As a Bayesian methods, BASTION provides uncertainty quantification via credible region. STR, a Frequentist counterpart, is the only other decomposition method with uncertainty quantification. We compare empirical coverage and average interval width to assess the accuracy and precision of uncertainty quantification provided by BASTION's credible intervals and STR's confidence intervals.

Models are implemented in \texttt{R} \cite{base}. TBATS and MSTL are implemented using the \texttt{forecast} package \cite{forecast_pack}, and STR is implemented using the \texttt{stR} package \cite{str_pack}. For BASTION, we extend the efficient sampling approach for the Gaussian state-space model framework from \cite{rue}, and the stochastic volatility model from \cite{stochvol}.
\subsection{Results}
\begin{table}[ht]
\centering
\resizebox{0.9\linewidth}{!}{%
\begin{tabular}{lcccc}
\toprule
\multicolumn{1}{c}{ } & \multicolumn{4}{c}{\textbf{Mean Squared Error}} \\
\cmidrule(l{3pt}r{3pt}){2-5}
 & TBATS & MSTL & STR & BASTION \\
\midrule
\addlinespace[0.3em]
\multicolumn{5}{l}{\textbf{DGP 1}} \\
\hspace{1em}Signal       & 13.878 & 11.760 & 10.829 & \textbf{0.376} \\
\hspace{1em}Trend        & 3.624  & 13.227 & 13.785 & \textbf{0.581} \\
\hspace{1em}Seasonality  & 11.672 & 2.712  & 2.575  & \textbf{0.536} \\
\addlinespace[0.3em]
\multicolumn{5}{l}{\textbf{DGP 2}} \\
\hspace{1em}Signal       & 1.411  & 1.214  & 0.7020 & \textbf{0.3922} \\
\hspace{1em}Trend        & 0.621  & 0.184  & 0.0878 & \textbf{0.0579} \\
\hspace{1em}Seasonality  & 0.838  & 1.032  & 0.6156 & \textbf{0.3412} \\
\addlinespace[0.3em]
\multicolumn{5}{l}{\textbf{DGP 3}} \\
\hspace{1em}Signal       & 11.307 & 3.041  & 0.706  & \textbf{0.439} \\
\hspace{1em}Trend        & 10.509 & 0.364  & \textbf{0.274} & 0.293 \\
\hspace{1em}Seasonality  & 0.900  & 2.683  & 0.444  & \textbf{0.278} \\
\addlinespace[0.3em]
\multicolumn{5}{l}{\textbf{DGP 4}} \\
\hspace{1em}Signal       & 11.829 & 11.430 & 20.548 & \textbf{2.877} \\
\hspace{1em}Trend        & 11.111 & 12.328 & 13.431 & \textbf{5.210} \\
\hspace{1em}Seasonality  & 5.364  & 3.045  & 11.358 & \textbf{2.562} \\
\bottomrule
\end{tabular}}
\caption{Mean Squared Error (MSE) for Trend, Seasonality, and their combined component (Signal). Each data generating processes (DGP) is replicated 1000 times.}
\label{tab:mse_output}
\end{table}
Table~\ref{tab:mse_output} reports mean squared error (MSE) for BASTION, MSTL, TBATS, and STR. Across all four simulation scenarios, BASTION consistently achieves the lowest or near-lowest MSE for the trend, seasonality, and combined signal components. The improvement is most pronounced in DGP~4, which includes both heteroskedastic noise and outlier components. Because BASTION explicitly models these features unlike the competing methods, it is better able to adapt to the data generating process.

\begin{table}[ht]
\centering
\resizebox{0.95\linewidth}{!}{%
\begin{tabular}{lcccc}
\toprule
\multicolumn{1}{c}{ } 
& \multicolumn{2}{c}{\textbf{Empirical Coverage}} 
& \multicolumn{2}{c}{\textbf{Interval Width}} \\
\cmidrule(l{3pt}r{3pt}){2-3}
\cmidrule(l{3pt}r{3pt}){4-5}
 & STR & BASTION & STR & BASTION \\
\midrule
\addlinespace[0.3em]

\multicolumn{5}{l}{\textbf{DGP 1}} \\
\hspace{1em}Signal       & 0.799 & \textbf{0.998} & 4.048 & 5.306 \\
\hspace{1em}Trend        & 0.616 & \textbf{0.970} & 1.934 & 1.993 \\
\hspace{1em}Seasonality  & 0.815 & \textbf{0.999} & 2.114 & 3.314 \\

\addlinespace[0.3em]
\multicolumn{5}{l}{\textbf{DGP 2}} \\
\hspace{1em}Signal       & 0.869 & \textbf{0.989} & 2.179 & 2.779 \\
\hspace{1em}Trend        & 0.716 & \textbf{0.976} & 0.992 & 0.854 \\
\hspace{1em}Seasonality  & 0.793 & \textbf{0.956} & 1.187 & 1.925 \\

\addlinespace[0.3em]
\multicolumn{5}{l}{\textbf{DGP 3}} \\
\hspace{1em}Signal       & \textbf{0.940} & 0.995 & 3.017 & 4.589 \\
\hspace{1em}Trend        & 0.812 & \textbf{0.929} & 1.621 & 1.959 \\
\hspace{1em}Seasonality  & 0.847 & \textbf{0.981} & 1.396 & 2.630 \\

\addlinespace[0.3em]
\multicolumn{5}{l}{\textbf{DGP 4}} \\
\hspace{1em}Signal       & 0.679 & \textbf{0.981} & 5.148 & 5.606 \\
\hspace{1em}Trend        & 0.668 & \textbf{0.939} & 2.296 & 2.246 \\
\hspace{1em}Seasonality  & 0.623 & \textbf{0.999} & 2.852 & 3.360 \\

\addlinespace[0.3em]
\bottomrule
\end{tabular}}
\caption{Empirical Coverage (EC) and Interval Width for Trend, Seasonality, and their combined component (Signal). EC is calculated using 95\% confidence intervals for STR and 95\% credible intervals for BASTION.}
\label{tab:ec_output}
\end{table}
Table~\ref{tab:ec_output} reports empirical coverage and average interval width for BASTION and STR, the only two methods considered that provide uncertainty quantification. In addition to evaluating uncertainty for the combined signal, we assess how well each method quantifies uncertainty for the individual components. BASTION consistently maintains coverage above the nominal 95\% level across all four DGPs, not only for the combined signal but also for both the trend and seasonality components. In contrast, STR attains nominal coverage only for the combined signal in DGP3 and fails to do so in DGPs1, 2, and 4. Moreover, STR substantially undercovers for the trend and seasonality components across all four DGPs, leading to comparatively narrower interval widths across all DGPs for STR.

The simulation results show that BASTION reliably recovers trend and seasonal structure across a broad range of data generating processes. Its advantages are most pronounced in settings with structural breaks, heteroskedastic noise, and additive outliers, where existing methods degrade. Moreover, BASTION provides well-calibrated uncertainty quantification for both the combined signal and the individual components. These results motivate applying BASTION to real-world time series, where such features are common across various disciplines.

\section{Empirical Data Analysis}
In this section, we apply BASTION to two real-world datasets that exhibit complex seasonality patterns, abrupt changes, heteroskedastic noise or combination of these characteristics. The first dataset consists of monthly U.S. airline traffic from 2003 to 2023, obtained from Kaggle \citep{us_airline_traffic}, which highlights sharp declines in passenger volume during significant disruptions such as the COVID-19 pandemic. For the second dataset, we analyze the average daily electricity demand (in Megawatts per hour) for the state of New York, sourced from the New York Independent System Operator (NYISO) via the U.S. Energy Information Administration \citep{nyis_electricity}. This dataset captures complexity with its long-term trend, with both weekly and yearly seasonal patterns and heteroskedastic noise.
\subsection{U.S. Airline Traffic Data}
\begin{figure*}[ht]
    \centering 
    \subfigure[TBATS]{%
        \includegraphics[width=0.42\linewidth]{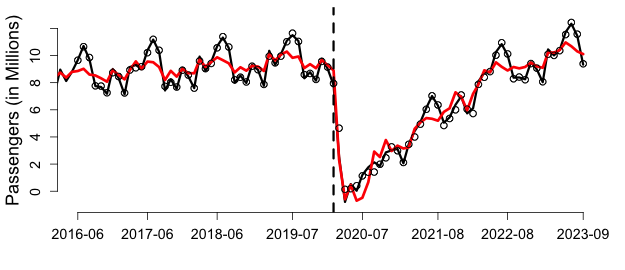}
        \label{fig:airline_t_tbats}
    }
    \subfigure[MSTL]{%
        \includegraphics[width=0.42\linewidth]{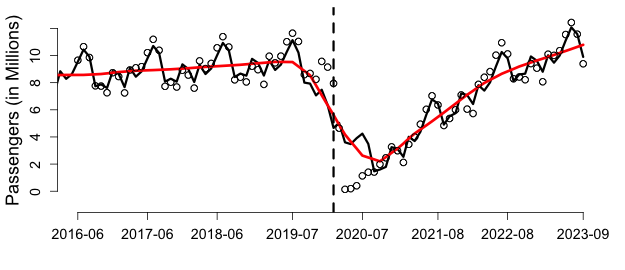}
        \label{fig:airline_t_mstl}
    }
    \subfigure[STR]{%
        \includegraphics[width=0.42\linewidth]{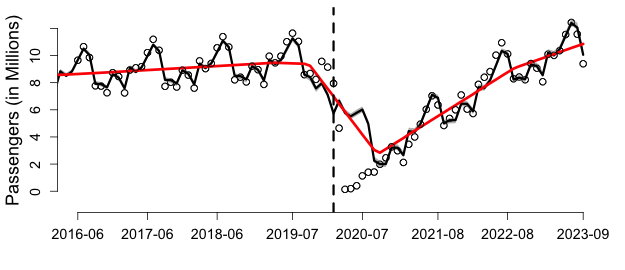}
        \label{fig:airline_t_STR}
    }
    \subfigure[BASTION]{%
        \includegraphics[width=0.42\linewidth]{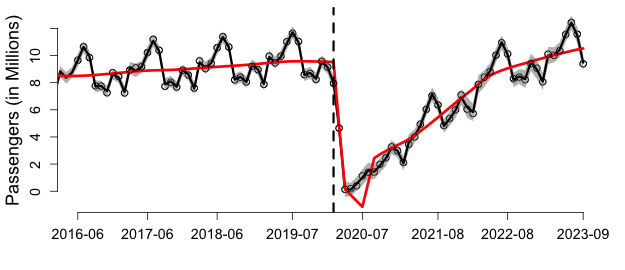}
        \label{fig:airline_t_BASTION}
    }
    \caption{Observed data (points); estimated trend + seasonality, ($\hat{\T} + \hat{\Sb}$), (black) with 95\% credible regions; and trend ($\hat{\T}$) component alone (red), for monthly U.S. international airline traffic, 2016 - 2023. The sharp decline in early 2020 reflects COVID-19 travel restrictions.}
    \label{fig:airline_comparison}
\end{figure*}
Monthly international airline traffic data from 2003 to 2023 (Figure~\ref{fig:airline_comparison}) exhibits a clear yearly seasonal pattern: passenger counts are lowest early in the year, rise steadily to a peak in July and August, decline sharply in the fall, and rebound moderately in December and January. As per the trend, we see a steady rise in the number of passenger with the significant drop in early 2020, corresponding to the COVID-19 pandemic and resulting travel restrictions, introducing a clear structural change in the trend. Figure~\ref{fig:airline_comparison} compares seasonality and trend estimates across TBATS, MSTL, STR, and BASTION. BASTION provides a smooth yet adaptive trend, capturing the abrupt COVID-19 drop while preserving overall structure. MSTL and STR oversmooth the break, treating it as gradual, whereas TBATS identifies the sharp change but yields noisy estimates. This example highlights BASTION's ability to maintain smoothness while adapting to abrupt changes. The full decomposition is provided in Appendix~\ref{App:addfigs}, Figure~\ref{fig:airline_BASTION}.

\subsection{New York Daily Average Electricity Demand}
\begin{figure}[ht]
    \centering 
     \includegraphics[width=\columnwidth]{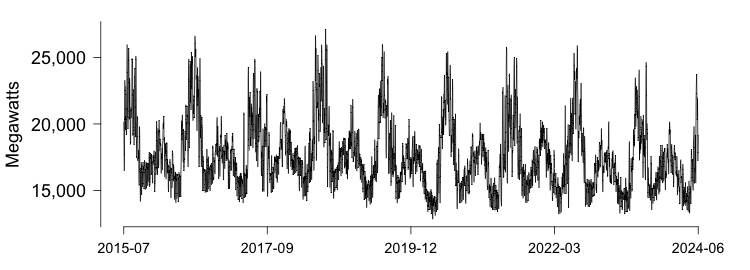}
    \caption{Daily average electricity demand in New York State from 2015-07-01 to 2024-06-30.}
   	\label{fig:elec_y}
\end{figure}
\begin{figure*}[!ht]
    \centering 
    \subfigure[TBATS]{%
        \includegraphics[width=0.48\linewidth]{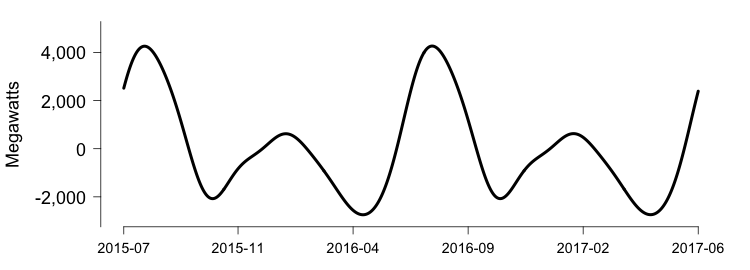}
        \label{fig:tbats_elec_sea}
    }
    \subfigure[MSTL]{%
        \includegraphics[width=0.48\linewidth]{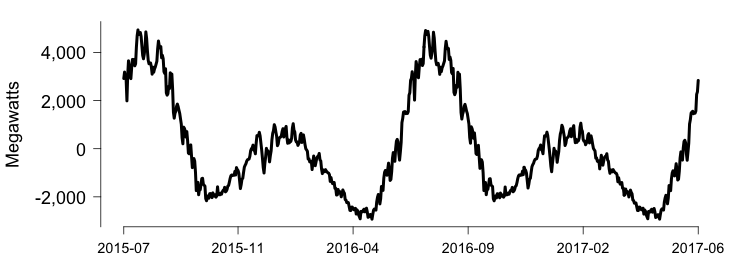}
        \label{fig:mstl_elec_sea}
    }
    \subfigure[STR]{%
        \includegraphics[width=0.48\linewidth]{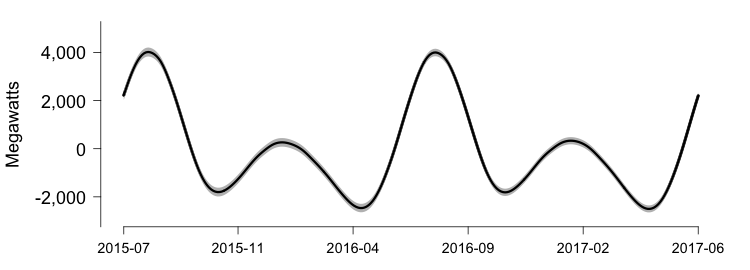}
        \label{fig:str_elec_sea}
    }
    \subfigure[BASTION]{%
        \includegraphics[width=0.48\linewidth]{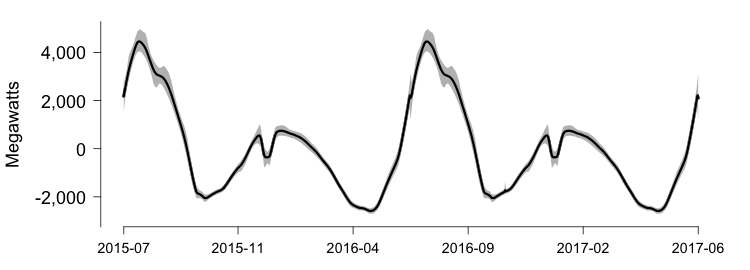}
        \label{fig:bastion_elec_sea}
    }
    \caption{Yearly seasonality estimates, ($\hat{\Sb}$), for daily average electricity demand in New York from July 2015 to June 2017. Figures (a) - (d) display estimates from TBATS, MSTL, STR, and BASTION, respectively.}
    \label{fig:elec_sea_comp}
\end{figure*}
The daily average electricity demand data for New York, spanning from July 1, 2015, to June 30, 2024, exhibits clear yearly seasonal patterns. As shown in Figure~\ref{fig:elec_y}, demand peaks in the summer months due to widespread air-conditioning use and increases again during winter, reflecting heating needs. Spring and fall, by contrast, experience milder temperatures and consequently lower demand.

As shown in Figure~\ref{fig:elec_sea_comp}, all four methods accurately capture the broad seasonal patterns typical of the East Coast, with peak electricity demand during the summer months of July and August due to increased cooling needs. This is followed by a steady decline in the fall and a moderate rise during the winter, reflecting increased heating requirements.

One interesting feature visible in the BASTION and MSTL estimates is a small dip in demand lasting approximately 30 days, from mid-December to mid-January, coinciding with the holiday slowdown in commercial and industrial activity. This reduction in electricity use drives the observed drop. BASTION distinguishes itself by detecting such subtle shifts while maintaining smooth, accurate seasonal estimates. The full BASTION decomposition, including trend and weekly seasonality, is provided in Appendix~\ref{App:addfigs}, Figure~\ref{fig:elec_decom}.

\begin{figure}[ht]
    \centering 
    \includegraphics[width=0.8\columnwidth]{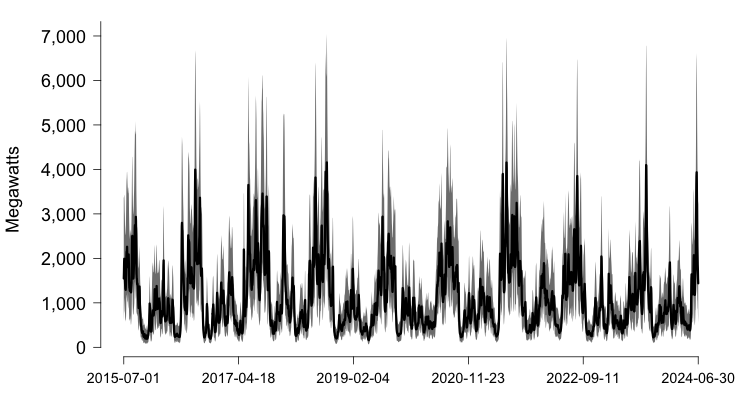}
    \caption{Volatility estimate based on BASTION for daily average electricity demand from 2015-07-01 to 2024-06-30. 95\% credible regions are drawn in dark grey.}
    \label{fig:elec_vol}
\end{figure}
Another defining characteristic of BASTION, that no other decomposition model provides, is its ability to explicitly estimate time varying volatility. BASTION's estimate of time-varying volatility, depicted in Figure~\ref{fig:elec_vol}, highlights not only the heteroskedasticity of the noise term but also the presence of seasonality in the volatility itself as well. Specifically, as the estimate suggests, higher volatility is observed during the winter and summer months, while lower volatility occurs during the spring and fall. This insight is particularly relevant for electricity management, where anticipating periods of elevated volatility informs resource allocation, capacity planning, and pricing strategies.
\section{Conclusion}
We introduced BASTION, a flexible Bayesian framework for decomposing time series into trend and multiple seasonalities while explicitly modeling outliers and volatility. Unlike existing methods, BASTION is able to capture abrupt changes in both the trend and seasonality component, while also providing rigorous uncertainty quantification. Simulation results demonstrates that BASTION outperforms TBATS, STR, and MSTL in both accuracy and empirical coverage in most cases. Empirical analyses of airline passenger data and electricity demand further validated its ability to capture complex temporal dynamics and provide interpretable insights. To facilitate further research and practical use, BASTION is available as an R package at \url{https://github.com/Jasoncho0914/BASTION}.

Several directions for future work remain. One natural extension is to evaluate how improved decomposition translates to downstream forecasting performance, particularly in settings with structural breaks and heteroskedastic noise. Another promising direction is to incorporate additional latent structure, such as Bayesian change point detection, to explicitly identify regime shifts while accounting for evolving seasonal patterns. These extensions would further enhance the flexibility and applicability of the BASTION framework.
\newpage
\subsubsection*{Acknowledgments}
\bibliographystyle{plainnat}
\bibliography{ref}

\newpage
\section*{Checklist}
\begin{enumerate}
  \item For all models and algorithms presented, check if you include:
  \begin{enumerate}
    \item A clear description of the mathematical setting, assumptions, algorithm, and/or model. \textbf{Yes}
    \item An analysis of the properties and complexity (time, space, sample size) of any algorithm. \textbf{Yes}
    \item (Optional) Anonymized source code, with specification of all dependencies, including external libraries. \textbf{Yes}
  \end{enumerate}

  \item For any theoretical claim, check if you include:
  \begin{enumerate}
    \item Statements of the full set of assumptions of all theoretical results. \textbf{Yes}
    \item Complete proofs of all theoretical results. \textbf{Yes}
    \item Clear explanations of any assumptions. \textbf{Yes}     
  \end{enumerate}

  \item For all figures and tables that present empirical results, check if you include:
  \begin{enumerate}
    \item The code, data, and instructions needed to reproduce the main experimental results (either in the supplemental material or as a URL). \textbf{Yes}
    \item All the training details (e.g., data splits, hyperparameters, how they were chosen). \textbf{Yes}
    \item A clear definition of the specific measure or statistics and error bars (e.g., with respect to the random seed after running experiments multiple times). \textbf{Yes}
    \item A description of the computing infrastructure used. (e.g., type of GPUs, internal cluster, or cloud provider). \textbf{Yes}
  \end{enumerate}

  \item If you are using existing assets (e.g., code, data, models) or curating/releasing new assets, check if you include:
  \begin{enumerate}
    \item Citations of the creator If your work uses existing assets. \textbf{Not Applicable}
    \item The license information of the assets, if applicable. \textbf{Not Applicable}
    \item New assets either in the supplemental material or as a URL, if applicable. \textbf{Not Applicable}
    \item Information about consent from data providers/curators. \textbf{Not Applicable}
    \item Discussion of sensible content if applicable, e.g., personally identifiable information or offensive content. \textbf{Not Applicable}
  \end{enumerate}

  \item If you used crowdsourcing or conducted research with human subjects, check if you include:
  \begin{enumerate}
    \item The full text of instructions given to participants and screenshots. \textbf{Not Applicable}
    \item Descriptions of potential participant risks, with links to Institutional Review Board (IRB) approvals if applicable. \textbf{Not Applicable}
    \item The estimated hourly wage paid to participants and the total amount spent on participant compensation. \textbf{Not Applicable}
  \end{enumerate}
\end{enumerate}

\clearpage
\appendix
\thispagestyle{empty}

\onecolumn
\aistatstitle{Supplementary Materials}
\section{PROOFS}
\subsection{Proof of Theorem 1}\label{pf:id}
Given a length $N$ time-series $\Y := \{Y_t\}_{t=1}^{N}$, the goal is to characterize the condition in which the following
\begin{align*}
	\hat{\T},\hat{\Sb_1},&\ldots,\hat{\Sb_P}  = \argmin_{\T,\Sb_1,\ldots,\Sb_P}\bigg\{ \|Y - \T - \sum_{i=1}^{P}\Sb_i\|^2_2 + \lambda_{0} \|D_{T}\T\|^2_2 +\sum_{i=1}^{P}\lambda_{i}\|D_{S,i}\Sb_i\|^2_2\bigg\},
\end{align*}
is uniquely identifiable. 

The first order condition of the objective function gives us the following sets of equations, 
\begin{align*}
		&A_{0}\hat{\T}  =  \Y - \sum_{i=1}^{P}\hat{\Sb_i}\\
		&A_{1}\hat{\Sb_1}  =  Y -\hat{\T} - \sum_{j \neq 1}\hat{\Sb_{j}}\\
		& \ldots \\
		&A_{P}\hat{\Sb_P}  =  Y -\hat{\T} - \sum_{j \neq P}\hat{\Sb_{j}},
\end{align*}
where $A_{0} = (I + \lambda_{0} L_{0})$, $L_{0} = (D'_{\T} D_{\T})$, and $A_{j} = (I + \lambda_{j} L_{j})$, $L_{j} = (D_{S,j}'D_{S,j})$ for $j  = 1,\ldots P$. This can be written as a linear system:
\begin{align*}
	\begin{bmatrix}
		A_{0} & I & \ldots & I \\
		I     & A_{1} & \ldots & I\\
		\vdots & \vdots & \ddots &\vdots \\
		I & I  & \ldots & A_{P}
	\end{bmatrix}
	\begin{bmatrix}
		\hat{\T} \\
		\hat{\Sb_1} \\
		\vdots \\
		\hat{\Sb_P}
	\end{bmatrix} = 
	\begin{bmatrix}
		\Y \\
		\Y \\
		\vdots \\
		\Y
	\end{bmatrix}.
\end{align*}
If the matrix $H = \begin{bmatrix}
		A_{0} & I & \ldots & I \\
		I     & A_{1} & \ldots & I\\
		\vdots & \vdots & \ddots &\vdots \\
		I & I  & \ldots & A_{P}
	\end{bmatrix}$, is full rank, the unique solution exists. This is equivalent to checking the dimension of the kernel, nullity, of the matrix. If the nullity is 0, $H$ is full ranked by the rank nullity relation.

First we show that $\ub = (u_{0},\ldots,u_{P}) \in \text{ker}(H)$ if and only if $\lambda_{j}L_{j}u_j = 0, \forall j \in \{0,\ldots,P\} $ and $\sum_{j=0}^{P} u_j = 0$.

($\Rightarrow$) Given $\ub := (u_0,u_1,\ldots,u_{P}) \in \text{ker}(H)$, define $s := \sum_{j=0}^{P} u_j$. By the assumption that $\ub \in \text{ker}(H)$, $H\ub = 0$. Or equivalently, for each block $\forall j \in \{0,\ldots,P\},$
\begin{equation}
	\begin{aligned}
			A_{j} u_j + \sum_{i \neq j}^{P} u_i = A_{j} u_j + \sum_{i \neq 0}^{P} u_i + (u_j - u_j) = (A_{j} - I)u_j  + \sum_{i=0}^{P}u_i = (A_{j} - I) u_j +s = (\lambda_{j}L_{j})u_j + s=  0		
	\end{aligned}\label{eq:t1}
\end{equation}
The only solution to $s = 0$ because $\sum_{j=0}^{P}\lambda_j u_j' L_j u_j = \sum_{j=0}^{P} -u_j's =  -(\sum_{j=0}^{P} u_j)'s = -\|s\|^2_2 > 0 $. 

($\Leftarrow$) Fix $\ub = (u_{0},\ldots,u_{P})$ such that $\lambda_{j}L_{j}u_j = 0, \forall j \in \{0,\ldots,P\} $ and $\sum_{j=0}^{P} u_j = 0$. We need to show that $H \ub = 0$, which can be verified by the same derivation shown in Equation~\ref{eq:t1} but backwards. $\forall j \in \{0,\ldots,P\},$
\begin{align*}
	&A_{j}u_j + \sum_{i\neq 0}^{P}u_i = (A_{j}-I)u_j + \sum_{i = 0}^{P}u_i = \lambda_jL_ju_j + \sum_{i=0}^{P}u_i = 0.
\end{align*}
Lastly, We argue that $\ker(L_0) = \ker(D_{T})$ and $\ker(L_j) = \ker(D_{S,j})$ for $j \geq 1. $  

$(\Rightarrow)$ If $u_0 \in \ker(L_{T})$, $L_T u_0$ = $D_{T}'D_{T}u_0 = 0$. Since positive semidefinite, $u_0'D_{T}'D_{T}u_0 = (D_{T}u_0)'D_{T}u_0  \geq 0$, which is only possible if and only if $D_{T}u_0 = 0$. The same argument applies to $D_{S,j}$ for $j \geq 1$.

$(\Leftarrow)$ if $u_0 \in \ker(D_{T})$, $D_{T}u_0 = 0$ thus, $(D_{T}u_0)'(D_{T}u_0) = u_0' L_T u_0 = 0$. Therefore, $L_{T}u_0 = 0$.

This proves that $\text{ker}(H)$ consists of vectors from each null space of the operators: $D_T D_{S,1},\ldots,D_{S,P}$, whose sum is zero. This can be characterized by $V := D_{T}\oplus_{j=1}^{P} \text{ker}(D_{S,j})$ and define the sum map $\Phi : V \rightarrow \mathbb{R}^N$, where $\Phi(v_{0},\ldots,v_{P}) = \sum_{j=0}^{P}v_{j}$. Therefore, $\text{ker}(H) = \text{ker}(\Phi)$. It follows that 
$$ \dim\ker(H) \;=\;\dim\ker(\Phi)
\;=\dim\ker(D_T) + \;\sum_{j=1}^P \dim\ker(D_j)\;-\;\dim\Big(\ker{D_{T}}+\sum_{j=1}^P \ker(D_j)\Big).$$
The matrix $H$ is uniquely identifiable if and only if the dimension of the null space is 0 by the rank-nullity relation.
\subsection{Proof of Corollary 1.1}\label{pf:c11}
Define $2 < k < N$, where $k$ is the length of the seasonality and $N$ be the length of the data. $\1 := (1,\ldots,1)^T \in \mathbb{R}^N$. $B$ is the back shift operator, $Bx_t = x_{t-1}$ for $t \geq 2$. For the trend $\ker(D_{T}) = \text{span}\{\1,m\}$, where $m = (1,2,\ldots N)'$. This is because $D_{T}x = 0$ if and only if differences vanish i.e., $x_t = a + bt$. Two linearly independent solutions are $\1$ and $m$. For the seasonal differencing operator, $\ker(D_{S,1})  = \{x \in \mathbb{R}^N: x_{t} = x_{t-k}, \forall t > k \}$. The first $k$ coordinates are free and determine all subsequent values. The shared null space is $\1$. 

For the seasonal recurrence $D_{S,1}$, each row has its first nonzero entry in a distinct column. For example row $r$ starts at column $r$, ending at $r+k$. therefore, rows are linearly independent and Rank of $D_{S,1}$ is $N - k$. By the rank nullity relation, the dimension of the null space is $k$. Note that $\1 \notin \ker(D_{S,1})$ and $m \notin \ker(D_{S,1})$. 
\subsection{Proof of Corollary 1.2}\label{pf:c12}
Let $P > 1$, $D_{T}$ be the second differencing operator and $D_{S,1},\ldots D_{S,P}$ be the seasonal differencing operator,$k_1 ,\ldots k_P$ be the length of the seasons for each seasonal components,
\begin{align*}
	&\dim(\ker(D_T))  + \sum_{j=1}^{P}\dim(\ker(D_{S,j}) - \dim(\ker(D_T) + U)\\ 
	&= 2 + \sum_{j=1}^{P}k_j - \dim(\ker(D_{T}) + U) \\ 
	&=2 + \dim(U)  + (\sum_{j=1}^{P}k_j  -\dim(U)) - \dim(\ker(D_{T}) + U)\\
	&= \dim(\ker(D_{T}) + U) + \sum_{j=1}^{P} k_j - \dim(U)
 \end{align*}

As shown previously $\ker(D_{T}) = \text{span}\{\1,m\}$, where $m = (1,2,\ldots N)'$, and $\ker(D_{S,j}) = \{x \in \mathbb{R}^N: x_{t} - x_{t-k_{j}},\forall t > k_j\}$, where $\dim\ker(D_{S,j}) = k_j$. $\dim(\ker(D_{T}) + U) = 1$.

For $U$ inclusion-exclusion gives us:
\begin{align*}
	\dim U = \sum_{j}\dim\ker(D_{s,j}) - \sum_{i<j}\dim\ker(D_{S,i} \cap D_{S,j}) + \sum_{i<j<l}\dim\ker(D_{S,i} \cap D_{S,j} \cap D_{S,l})- \dots + \dim\ker(\bigcap_{j=1}D_{S,j}).
\end{align*}
Note that the shared null space between two seasonal components  $D_{S,i}$ and $D_{S,j}$ is $gcd(k_i,k_j)$. 
\begin{align*}
	\dim U = \sum_{j}k_j - \sum_{i<j}\gcd(k_i,k_j) + \sum_{i<j<l}\gcd(k_i,k_j,k_l)- \dots + \gcd(k_1,\ldots k_P).
\end{align*}
Thus, the nullity of the design matrix $H$ in Theorem 1 reduces to 
\begin{align*}
	&\dim(\ker(D_T))  + \sum_{j=1}^{P}\dim(\ker(D_{S,j}) - \dim(\ker(D_T) + U) \\
	& = 1 + \sum_{i<j}\gcd(k_i,k_j) - \sum_{i<j<l}\gcd(k_i,k_j,k_l)+ \dots - \gcd(k_1,\ldots k_P)	
\end{align*}

If we assume $D_{S,1},\ldots D_{S,P}$ to be a recurrent operator,  the nullity is reduced by 1 as the null space between the seasonality and the trend does not interact.
 \begin{align*}
	&\dim(\ker(D_T))  + \sum_{j=1}^{P}\dim(\ker(D_{S,j}) - \dim(\ker(D_T) + U) \\
	& = \sum_{i<j}\gcd(k_i,k_j) - \sum_{i<j<l}\gcd(k_i,k_j,k_l)+ \dots - \gcd(k_1,\ldots k_P)	
\end{align*}

\subsection{Proof of Theorem 2}\label{pf:idmult}
Given a length $N$ time-series $\Y := \{Y_t\}_{t=1}^{N}$, we consider the following objective function:
\begin{align*}
	\hat{\T},\hat{\Sb_1},&\ldots,\hat{\Sb_P}  = \argmin_{\T,\Sb_1,\ldots,\Sb_P}\bigg\{\|Y - \T - \sum_{i=1}^{P}\Sb_i\|^2_2 + \sum_{m=1}^{q_T}\lambda^m_{0} \|D^m_{T}\T\|^2_2 + \sum_{m=1}^{q_S}\lambda^{m}_{i}\sum_{i=1}^{P}\|D_{S,i}^m\Sb_i\|^2_2 \bigg \},
\end{align*}
Here, we have $q_T$ and $q_S$ now represent the number of penalties for the trend and seasonal component respectively. The hyperparameter $\lambda_0^m$ for the trend and $\lambda_{i}^m$ is strictly positive scalar controlling the degree of penalty applied to each penalty operator. 

We provide a similar identifiability condition as in Theorem 1. Define $V_0 = \bigcap_{m=1}^{q_T} \text{ker}(L_0^m)$ and $V_{j} = \bigcap_{m=1}^{q_S}\ker(L_j^m)$ for $j >0$. In this proof, we show that we get a uniquely identifiable solution if:
$$ \sum_{j=0}^P \dim\ker(V_j)\;-\;\dim\Big(\sum_{j=0}^P \ker(V_j)\Big) = 0 $$

The same first order condition therefore the same system of equation in Theorem 1 holds, but $A_{0},\ldots A_{P}$ are defined differently: $A_{0} = (I + \sum_{m=1}^{q_T} \lambda_{0}^m L_{0}^m)$, $L_{0}^m = (D^m_{\T})' D_{\T}^m$, for $m = 1,\ldots,q_T$ and $A_{j} = (I + \sum_{m=1}^{q_S}\lambda_{j}^m L_{j}^m)$, $L_{j}^m = (D_{S,j}^m)'D_{S,j}^m$, for $j  = 1,\ldots P$ and $m = 1 ,\ldots q_S$. 
\begin{align*}
	\begin{bmatrix}
		A_{0} & I & \ldots & I \\
		I     & A_{1} & \ldots & I\\
		\vdots & \vdots & \ddots &\vdots \\
		I & I  & \ldots & A_{P}
	\end{bmatrix}
	\begin{bmatrix}
		\hat{\T} \\
		\hat{\Sb_1} \\
		\vdots \\
		\hat{\Sb_P}
	\end{bmatrix} = 
	\begin{bmatrix}
		\Y \\
		\Y \\
		\vdots \\
		\Y
	\end{bmatrix}.
\end{align*}
Given $H = 	\begin{bmatrix}
		A_{0} & I & \ldots & I \\
		I     & A_{1} & \ldots & I\\
		\vdots & \vdots & \ddots &\vdots \\
		I & I  & \ldots & A_{P}
	\end{bmatrix}$, its nullity is considered. 
	
We show that $\ub = (u_0,\ldots,u_P) \in \ker(H)$ if and only if 1) $u_{0}$ is in the kernel of $(\sum_{m=1}^{q_T}\lambda_{0}^mL_{0}^m)$ and $u_j$ is in the kernel of $(\sum_{m=1}^{q_s}\lambda_{j}^m L_{j}^m)$ for $j = 1,\ldots, P$ , and $\sum_{j=0}^Pu_j = 0$. 

($\Rightarrow$) Let $\ub  = (u_0,\ldots,u_P) \in \ker(H)$, and define $ s:= \sum_{j=0}^{P}u_{j}$. For the trend component:
\begin{equation*}
	\begin{aligned}
			A_{0} u_0 + \sum_{i \neq 0}^{P} u_i = (A_{0} - I)u_0  + \sum_{i=0}^{P}u_i = (A_{0} - I) u_0 + s = \sum_{m=1}^{q_T}\lambda_{0}^mL_{0}^m u_0 + s= \bigg(\sum_{m=1}^{q_T}\lambda_{0}^mL_{0}^m\bigg) u_0 + s		
	\end{aligned}
\end{equation*}
and for the seasonal components, $j \in 1,\ldots P$,
\begin{equation*}
	\begin{aligned}
			A_{j} u_j + \sum_{i \neq j}^{P} u_j = \bigg(\sum_{m=1}^{q_s}\lambda_{j}^m L_{j}^m\bigg)u_j + s=  0.
	\end{aligned}
\end{equation*}
We also show that only possible $s = 0$. We use the fact that $L_j^m$ are positive semi-definite and $\lambda_{j}^m>0$:
\begin{align*}
	&u'_0(\lambda_{0}^mL_{0}^m) u_0 + \sum_{j=1}^{P}\sum_{m=1}^{q_s}u_{j}'(\lambda_{j}^m L_{j})u_j \geq 0 \\  
	&u'_0(\lambda_{0}^mL_{0}^m) u_0 + \sum_{j=1}^{P}u_{j}'\sum_{m=1}^{q_s}(\lambda_{j}^m L_{j})u_j \geq 0 \\
	&u'_0(-s)+ \sum_{j=1}^{P}u_{j}'(-s) = (\sum_{j=0}^{P}u_{j})'(-s) = -\|s\|^2_2\geq 0.
\end{align*}

($\Leftarrow$)  Fix $\ub = (u_0,\ldots,u_P)$ such that $u_{0}$ is in the kernel of $(\sum_{m=1}^{q_T}\lambda_{0}^mL_{0}^m)$ and $u_j$ is in the kernel of $(\sum_{m=1}^{q_s}\lambda_{j}^m L_{j}^m)$ for $j = 1,\ldots, P$ , and $\sum_{j}^Pu_j = 0$. For $j = 0$,
\begin{align*}
	&A_{0}u_0 + \sum_{i\neq 0}^{P}u_i = (A_{0}-I)u_0 + \sum_{i = 0}^{P}u_i = \bigg(\sum_{m=1}^{q_T}\lambda_0^mL_0^m\bigg)u_0 + \sum_{i=0}^{P}u_i = 0.
\end{align*}

This proves that $\ub = (u_0,\ldots,u_P) \in \ker(H)$ if and only if 1) $u_{0}$ is in the kernel of $(\sum_{m=1}^{q_T}\lambda_{0}^mL_{0}^m)$ and $u_j$ is in the kernel of $(\sum_{m=1}^{q_s}\lambda_{j}^m L_{j}^m)$ for $j = 1,\ldots, P$ , and $\sum_{j}^Pu_j = 0$. 

Lastly, $\ker(\sum_{m=1}^{q_T} \lambda_0^m L_{0}^m) = \bigcap _{m=0}^{q_T}\ker(L_0^m)$ and $\ker(\sum_{m=1}^{q_S} \lambda_j^m L_{j}^m) = \bigcap _{m=0}^{q_S}\ker(L_j^m)$ for $j > 1$. 

($\Rightarrow$) Let's assume $u_0 \in \text{ker}(\sum_{m=1}^{q_T} \lambda_0^m L_{0}^m)$. Because $L_0^m$ is positive semi-definite, $u_0' \lambda_0^mL_0^m u_0 \geq 0$, $\forall m$. This requires that $u_0' \lambda_0^mL_0^m u_0 = 0$, $\forall m$. Thus $u_0 \in \bigcap_{m=1}^{q_T}\ker(L_0^m)$. The same argument can be made to show that $u_j \in \bigcap_{m=1}^{q_S}\ker(L_j^m) \forall j>1$.

($\Leftarrow$) If $u_0  \in \bigcap_{m=1}^{q_T}\ker(L_0^m)$, then $u_0' \lambda_0^mL_0^m u_0 = 0$ for all $m$, thus $\sum_{m=1}^{q_T}u_0' \lambda_0^mL_0^m u_0 = 0.$ $u_0 \in \ker(\sum_{m=1}^{q_T}\lambda_{0}^mL_{0}^m)$. The same argument applies to show $u_{j} \in \ker(\sum_{m=1}^{q_S}\lambda_{j}^mL_{j}^m)$ for $j>0$. 

Define $V_0 = \bigcap_{m=1}^{q_T} \text{ker}(L_0^m)$ and $V_{j} = \bigcap_{m=1}^{q_S}\ker(L_j^m)$ for $j >0$. 

This proves that $\text{ker}(H)$ consists of vectors from each null space of $V_j$, whose sum is zero:
\[\text{ker}(H) = \{(u_0,\ldots,u_P)|\forall j, u_{j} \in V_j , \sum_{j=0}^{P}u_{j} = 0\}.\]
The condition above can be characterized by $V := \oplus_{j=0}^{P} V_j$ and define the sum map $\Phi : V \rightarrow \mathbb{R}^N$, where $\Phi(v_{0},\ldots,v_{P}) = \sum_{j=0}^{P}v_{j}$. Therefore, $\text{ker}(H) = \text{ker}(\Phi)$. It follows that 
$$ \dim\ker(H) = \sum_{j=0}^P \dim\ker(V_j)\;-\;\dim\Big(\sum_{j=0}^P \ker(V_j)\Big).$$

By the nullity-rank relationship, the matrix is full-rank if and only if above equals to 0.
\subsection{Proof of Corollary 2.1}\label{pf:c21}
Using the same notation as in Theorem~2, we first note that 
\[
V_0 = \ker(D_T) = \mathrm{span}\{\mathbf 1,m\}, 
\quad m = (1,2,\ldots,N)^\top,
\]
$\dim V_0=2$. For each seasonal component $S_j$, the joint kernel of the second and seasonal differencing operators is
\[
V_j = \ker(D_{S,j}^1)\cap \ker(D_{S,j}^2) = \mathrm{span}\{\mathbf 1\},
\]
following from Corollary 1.1 (Appendix~\ref{pf:c11}).
Imposing the centering constraint $\mathbf 1^\top S_j=0$ eliminates this direction, so in fact $V_j=\{0\}$ for all $j\ge1$. Hence
\[
\sum_{j=0}^P \dim V_j = \dim V_0 = 2.
\]
Moreover,
\[
\sum_{j=0}^P V_j = V_0,
\]
which also has dimension $2$. Thus the condition of Theorem~2 holds with equality, and the decomposition is identifiable.

If we assume $D_{S_j}^2$ is the seasonal recurrence operator instead of the differencing operator
$$ V_{j} = \ker(D_{S,j}^1) \cap \ker(D_{S,j}^2) = \{0\},$$
as shown in Corollary~\ref{pf:c11}. Therefore, the decomposition becomes identifiable.

\section{Full Conditional Distribution for Gibbs Sampling}\label{App:gibbs}
\subsection{Full Model}\label{sec:Full Model}
The exact parameterization and the prior distributions of BASTION used for both the simulation and empirical study section is described below:
\begin{align*}
    &\textbf{Observation Equation} & &  \\
    &y_{t} =  T_{t} + \sum_{i=1}^{P}S_{i,t} + \zeta_t + R_{t}, & & [R_{t}|\sigma_{y},\nu_{t}] \stackrel{iid}{\sim} N(0,\sigma_y^2\nu_{t}^2) \\
    & & & \\
    &\textbf{Trend} & & \\ 
	&[T_{t}|\sigma_{y},\eta_{T,t}] \sim N(0, \sigma_y^2 \eta_{T,t}^2) & &\forall t \in \{1,2\}, \\
	&[\Delta^2 T_{t}|\sigma_{y},\tau_{T},\eta_{T,t}] \sim N(0, \sigma_y^2 \tau_{T}^2 \eta_{T,t}^2)& & t \geq 3,\\
	& [\tau_{T}] \sim C^+(0,1/N), & & [\eta_{T,t}] \stackrel{iid}{\sim} C^+(0,1). \\
    & & & \\
    &\textbf{Multiple Seasonality} & &  \\ 
	&[S_{i,1}] = 0, & & [S_{i,2} | \sigma_{y}, \eta_{S_i,2}] \sim N(0, \sigma_y^2 \eta_{S_i,2}^2),\\
    &[\Delta^2 S_{i,t} | \sigma_y, \tau_{S_i}, \eta_{S_i,t} ] \sim N(0, \sigma_y^2 \tau_{S_i}^2 \eta_{S_i,t}^2) & & \forall t \in \{3, \ldots, k_i\},\\
    &[(1 - B)^{k_i} S_{i,t} | \sigma_{y}, \tau_{S_i}, \eta_{S_i,t}] \sim N(0, \sigma_y^2 \tau_{S_i}^2 \eta_{S_i,t}^2) & & \forall t \geq (k_i+1),\\
    &[\tau_{S_i}] \sim C^+(0,1/N), & & [\eta_{S_i,t}] \stackrel{iid}{\sim} C^+(0,1).\\
    & & & \\
    &\textbf{Outliers} & & \\ 
    &[\zeta_{t}|\sigma_{y},\eta_{\zeta,t}] \sim N(0, \sigma_y^2 \eta^2_{\zeta,t}), & & [\eta_{\zeta,t}|\tau_{\zeta},\xi_{\zeta,t}] \sim C^+(0, \tau_{\zeta} \xi_{\zeta,t}) \\
    &[\tau_{\zeta}] \sim C^+(0,1), & & [\xi_{\zeta,t}] \stackrel{iid}{\sim} C^+(0,1).\\
    & & & \\
    &\textbf{Remainder} & & \\ 
    &[\sigma^2_y] \sim \sigma^{-2}_y d\sigma^2_y & & \\
    &\log(\nu^2_t) = \mu + \phi(\log(\nu^2_{t-1}) - \mu) + \sigma_{\nu}\epsilon_{t}, & & [\epsilon_{t}]\stackrel{iid}{\sim} N(0,1) \\
    &[\mu] \sim N(0,100), & & [\phi] \sim \text{Beta}(5,1.5) \\
    &[\sigma^2_\nu] \sim IG(1/2,1/2) & & \\ 
 \end{align*}
Let's first define $\boldsymbol{Y} = [y_{1},\ldots,y_{N}]'$ and similarly for the variables $\boldsymbol{T}, \boldsymbol{S_i}, \boldsymbol{\zeta}, \boldsymbol{\nu}$. The full conditional posterior distributions for Gibbs sampling are derived in the following subsections.
\subsection{Trend}\label{sec:Trend}
For the likelihood, we have:
$$    [\boldsymbol{Y}|\boldsymbol{T},\ldots ]\sim N\bigg(\boldsymbol{T} + \sum_{i=1}^{P}\boldsymbol{S_{i}} +\bm{\zeta}, \sigma^2_{y}\boldsymbol{\nu^2}I \bigg).
$$
The prior distribution on $\boldsymbol{T}$ are imposed on its second differencing.
\begin{align*}
	D_T &= \begin{bmatrix}
    1 & 0 & 0 & 0 & \cdots & 0 \\
    0 & 1 & 0 & 0 & \cdots & 0 \\
    1 & -1 & 1 & 0 & \cdots & 0 \\
    0 & 0 & -1 & 1 & \cdots & 0 \\
    \vdots & \vdots & \vdots & \vdots & \ddots & \vdots \\
    0 & 0 & 0 & \cdots & -1 & 1 \\
\end{bmatrix}
\begin{bmatrix}
    1 & 0 & 0 & 0 & \cdots & 0 \\
    0 & 1 & 0 & 0 & \cdots & 0 \\
    0 & -1 & 1 & 0 & \cdots & 0 \\
    0 & 0 & -1 & 1 & \cdots & 0 \\
    \vdots & \vdots & \vdots & \vdots & \ddots & \vdots \\
    0 & 0 & 0 & \cdots & -1 & 1 \\
\end{bmatrix} = 
\begin{bmatrix}
    1 & 0 & 0 & \cdots & 0 & 0 & 0 \\
    0 & 1 & 0 & \cdots & 0 & 0 & 0 \\
    1 & -2 & 1 & \cdots & 0 & 0 & 0 \\
    0 & 1 & -2 & 1 & 0 & 0 & 0 \\
    \vdots & \vdots & \ddots & \ddots & \ddots & \vdots & \vdots \\
    0 & 0 & 0 & \ddots & -2 & 1 & 0 \\
    0 & 0 & 0 & \cdots & 1 & -2 & 1 \\
\end{bmatrix}.
\end{align*}
Thus, given $\boldsymbol{\sigma^2_{T}} := [\sigma^2_y\eta^2_{T,1},\sigma^2_y\eta^2_{T,2},\sigma^2_y\tau^2_{T}\eta^2_{T,3},\ldots, \sigma^2_y\tau^2_{T}\eta^2_{T,N}]' $
\begin{align*}
	[\boldsymbol{T}|\ldots ]&\sim N(0, A\boldsymbol{\sigma^2_{T}}IA') \\ 
	&\sim N\bigg(0, \bigg(D_T'\boldsymbol{\frac{1}{\sigma^2_{T}}}I D_T\bigg)^{-1} \bigg).  
\end{align*}
Let's define $Q_{T} := (D_T)'\boldsymbol{\frac{1}{\sigma^2_{T}}}I(D_T)$. 
\begin{align*} 
    &Q_{T} = \frac{1}{\sigma_y^2\tau^2_{T}} \\
    &\left[\begin{smallmatrix}
         \bigg(\frac{\tau^2_{T}}{\eta^2_{T,1}} + \frac{1}{\eta^2_{T,3}}\bigg) & -\frac{2}{\eta^2_{T,3}} & -\frac{1}{\eta^2_{T,3}}  & 0 & \ldots & \ldots & 0\\
         -\frac{2}{\eta^2_{T,3}}  & \bigg(\frac{\tau^2_{T}}{\eta^2_{T,2}} + \frac{4}{\eta^2_{T,3}} + \frac{1}{\eta^2_{T,4}}\bigg) & \frac{-2}{\eta^2_{T,3}} - \frac{2}{\eta^2_{T,4}} &  -\frac{1}{\eta^2_{T,4}} &\ddots & \ddots & \vdots\\
         -\frac{1}{\eta^2_{T,3}} & \frac{-2}{\eta^2_{T,3}} - \frac{2}{\eta^2_{T,4}}  & \bigg(\frac{1}{\eta^2_{T,3}} + \frac{4}{\eta^2_{T,4}} + \frac{1}{\eta^2_{T,5}}\bigg)  & \frac{-2}{\eta^2_{T,4}} - \frac{2}{\eta^2_{T,5}} &  -\frac{1}{\eta^2_{T,5}} & \ddots &\ddots \\
         0 & \ddots  & \ddots & \ddots & \ddots & \ddots & 0\\
         \vdots       &\ddots   & \ddots & \ddots  & \bigg(\frac{1}{\eta^2_{T,N-2}} + \frac{4}{\eta^2_{T,N-1}}  + \frac{1}{\eta^2_{T,N}}\bigg) & \bigg(\frac{-2}{\eta^2_{T,N-1}}-\frac{2}{\eta^2_{T,N}}\bigg) & -\frac{1}{\eta^2_{T,N}}\\
         \vdots       &\ddots   & \ddots & \ddots  & \bigg(\frac{-2}{\eta^2_{T,N-1}}-\frac{2}{\eta^2_{T,N}}\bigg) & \bigg(\frac{1}{\eta^2_{T,N-1}} + \frac{4}{\eta^2_{T,N}} \bigg) & -\frac{2}{\eta^2_{T,N}}\\
         0 & \ldots & \ldots & \ldots & -\frac{1}{\eta^2_{T,N}} & -\frac{2}{\eta^2_{T,N}} & \frac{1}{\eta^2_{T,N}} \\
    \end{smallmatrix}\right]
\end{align*}
Therefore,
\begin{align*}
	[\boldsymbol{T}|\boldsymbol{y},\ldots ]\sim N\bigg( \bigg(Q_{T}+ \frac{1}{\sigma^2_{y}\boldsymbol{\nu^2}}I\bigg)^{-1} \bigg(\frac{\boldsymbol{Y}-\sum_{i=1}^{P}S_{i}-\bm{\zeta}}{\sigma^2_{y}\boldsymbol{\nu^2}}\bigg),\bigg(Q_{T}+ \frac{1}{\sigma^2_{y}\boldsymbol{\nu^2}}I\bigg)^{-1}\bigg).
\end{align*}
For the global parameter $\tau_{T}$ and the local parameter $\eta_{T,t}$ both following the half-Cauchy distribution, the parameter expansion as described in \cite{easyHSsampler} are used. Thus,
\begin{align*}
	& [\tau_{T}|\psi_{\tau_T},\ldots ]\sim IG(1/2,1/\psi_{\tau_T}) & [\psi_{\tau_T}] \sim IG(1/2,1)\\
	& [\eta_{T,t}|\psi_{\eta_{T},t}\ldots] \sim IG(1/2,1/\psi_{\eta_{T},t}) & [\psi_{\eta_{T},t}] \sim IG(1/2,1)
\end{align*}
Since they are conjugate priors, the conditional prior distributions are as follows:
\begin{align*}
	[\tau^2_{T}|\boldsymbol{Y},\psi_{\tau_T},\ldots ]&\sim IG\bigg(\frac{1}{2}+\frac{N-2}{2},\frac{1}{\psi_{\tau_T}} + \frac{1}{2\sigma^2_{y}}\sum_{t=3}^{N} \bigg(\frac{\Delta^2 T_{t}}{\eta_{T,t}}\bigg)^2 \bigg). \\ 
	[\psi_{\tau_T}|\boldsymbol{Y},\tau_{T},\ldots ]&\sim IG\bigg(1, 1+\frac{1}{\tau^2_{T}}\bigg). \\ 
	[\eta^2_{T,t}|\boldsymbol{Y},\psi_{\eta_{T},t}\ldots ]&\sim IG\bigg(1,\frac{1}{\psi_{\eta_T,t}} + \frac{1}{2}\bigg(\frac{T_{t}}{\sigma_{y}}\bigg)^2 \bigg), & t \in \{1,2\}. \\
	&\sim IG\bigg(1,\frac{1}{\psi_{\eta_T,t}} + \frac{1}{2}\bigg(\frac{\Delta^2 T_{t}}{\sigma_{y}\tau_{T}}\bigg)^2 \bigg), & t \geq 3. \\ 
	[\psi_{\eta_T,t}|\boldsymbol{Y},\eta_{T,t},\ldots] &\sim IG\bigg(1, 1+\frac{1}{\eta^2_{T,t}}\bigg). \\
\end{align*}

\subsection{Seasonality}
Let's fix $j \in \{1,\ldots,P\}$, and consider the seasonality term $\boldsymbol{S_{j}}$. Let $k_j$ denote the length of the cycle and to simplify notation let $\Sb := \Sb_j$. The seasonality parameter is penalized by both the seasonal differencing for the first $k$ component and are also penalized by the seasonal differencing operator:
\[
D_S = (\begin{bmatrix}
    I_{k-1} & 0 \\
    B_{3} & B_{2}
\end{bmatrix}
\begin{bmatrix}
    B_{1} & 0 \\
    0 & I_{N-k},
\end{bmatrix})^{-1}
\]
where each matrix \( B_{k,1} \), \( B_{k,2} \) has the following structure:
\begin{align*}
  &B_{1} = \begin{bmatrix}
    1 & 0 & 0 & \cdots & 0 \\
    2 & 1 & 0 & \cdots & 0 \\
    3 & 2 & 1 & \cdots & 0 \\
    \vdots & \vdots & \vdots & \ddots & \vdots \\
    k-1 & k-2 & k-3 & \cdots & 1 \\
\end{bmatrix}   
 \qquad
   B_{2} = \begin{bmatrix}
       I_{k} & 0 & 0 & \cdots & 0 \\
       I_{k} & I_{k} & 0 & \cdots & 0 \\
       I_{k} & I_{k} & I_{k} & \cdots & 0 \\
       \vdots & \vdots & \vdots & \ddots & \vdots \\
       I_{k} & I_{k} & I_{k} & \cdots & I_{k}
   \end{bmatrix}
 \qquad
   B_{3} = \begin{bmatrix}
       \bm{0}\\
       I_{k-1}  \\
       \bm{0}  \\
       I_{k-1}  \\
	   \bm{0}  \\
       \vdots
   \end{bmatrix}.
\end{align*}
$B_{1}$ is a \((k-1) \times (k-1)\) lower triangle matrix for un-differencing the second differencing operator on the first $k-1$ observations. $B_{2}$ is a \((N-k) \times (N-k)\)  lower triangle matrix and $B_{3}$ is a \((N-k) \times (k-1)\) matrix for un-differencing the seasonal differencing operator on the last $T-k$ observations. Similar to the derivation of $Q_{T}$ in Section~\ref{sec:Trend}, $Q_{S}^{-1} := \bigg((D_S)'\boldsymbol{\frac{1}{\sigma^2_{S}}}I D_S\bigg)$, where $\boldsymbol{\sigma^2_{S}} := [\sigma^2_y\eta^2_{S,2},\sigma^2_y\tau^2_{S}\eta^2_{S,3},\sigma^2_y\tau^2_{S}\eta^2_{S,4},\ldots, \sigma^2_y\tau^2_{S}\eta^2_{S,N}]'$. The precision  matrix $Q_{S}$ has a block-tridiagonal structure with blocks separated by the seasonality k except for the top left \((k-1) \times (k-1)\) submatrix, which has a tridiagonal structure resulting from the second differencing.

The conditional posterior distribution for $\boldsymbol{S}$ therefore, 
\begin{align*}
	[\boldsymbol{S}|\boldsymbol{y},\ldots ]  \sim N\bigg( \bigg(Q_{S}+ \frac{1}{\sigma^2_{y}\bm{\nu}^2}I\bigg)^{-1} \bigg(\frac{\boldsymbol{Y}- \boldsymbol{T} -\sum_{i \neq j} \boldsymbol{S_{i}}- \bm{\zeta}}{\sigma^2_{y}\bm{\nu}^2}\bigg),
	\bigg(Q_{S}+ \frac{1}{\sigma^2_{y}\bm{\nu}^2}I\bigg)^{-1}\bigg).
\end{align*}

The conditional posterior distributions of $\tau_{S}$ and $\bm{\eta_{S}}$ are identical to $\tau_{T}$ and $\bm{\eta_{T}}$ explored in Section~\ref{sec:Trend}. With the parameter expansion of the horseshoe prior:
\begin{align*}
	[\tau^2_{S}|\boldsymbol{Y},\psi_{\tau_S},\ldots ]&\sim IG\bigg(\frac{1}{2}+\frac{N-2}{2},\frac{1}{\psi_{\tau_S}} + \frac{1}{2\sigma^2_{y}}\bigg(\sum_{t=3}^{k} \bigg(\frac{\Delta^2 S_{t}}{\eta_{T,t}}\bigg)^2 &+ \sum_{t=(k+1)}^{N} \bigg(\frac{(1-B^{k})S_{t}}{\eta_{T,t}}\bigg)^2 \bigg)\bigg).  \\ 
	[\psi_{\tau_S}|\boldsymbol{Y},\tau_S,\ldots ]&\sim IG\bigg(1, 1+\frac{1}{\tau^2_{S}}\bigg). &\\ 
	[\eta^2_{S,t}|\boldsymbol{Y},\psi_{\eta_{S},t}\ldots] &\sim IG\bigg(1,\frac{1}{\psi_{\eta_S,t}} + \frac{1}{2}\bigg(\frac{S_{t}}{\sigma_{y}}\bigg)^2 \bigg), & t = 2, \\
	&\sim IG\bigg(1,\frac{1}{\psi_{\eta_S,t}} + \frac{1}{2}\bigg(\frac{\Delta^2 S_{t}}{\sigma_{y}\tau_{S}}\bigg)^2 \bigg),& t = 3, \ldots, k, \\ 
	&\sim IG\bigg(1,\frac{1}{\psi_{\eta_S,t}} + \frac{1}{2}\bigg(\frac{(1-B^{k})S_{t}}{\sigma_{y}\tau_{S}}\bigg)^2 \bigg),& t = (k+1), \ldots, N. \\ 
	[\psi_{\eta_S,t}|\boldsymbol{Y},\eta_{S,t},\ldots] &\sim IG\bigg(1, 1+\frac{1}{\eta^2_{S,t}}\bigg). & 
\end{align*}
\subsection{Outliers}
For the additive outlier term $\zeta_{t}$, we use the horseshoe+ prior by \cite{horseshoe_P} to provide more shrinkage than the one induced by the horseshoe prior. The parameter expansion by \cite{easyHSsampler} can be applied to the horseshoe+ prior:
\begin{align*}
	&[\zeta_{t}|\sigma_{y},\eta_{\zeta,t}] \sim N(0, \sigma^2_{y} \eta_{\zeta,t}^2), & [\eta^2_{\zeta,t}|\psi_{\eta_{\zeta},t}] \sim IG(1/2, 1/\psi_{\eta_{\zeta},t} ),\\ 
	&[\psi_{\eta_{\zeta},t}|\tau_{\zeta},\xi_{\zeta,t}] \sim IG(1/2, 1/(\tau^2_{\zeta}\xi^2_{\zeta,t})), &\\
	&[\tau^2_{\zeta}|\psi_{\tau_{\zeta}}] \sim IG(1/2, 1/\psi_{\tau_{\zeta}}), & 
	[\psi_{\tau_{\zeta}}] \sim IG(1/2, 1) , \\ 
	&[\xi^2_{\zeta,t}|\psi_{\xi_{\zeta},t}] \sim IG(1/2,1/\psi_{\xi_{\zeta},t}), & [\psi_{\xi_{\zeta},t}] \sim IG(1/2,1).
\end{align*}
Thus, we have the following conditional posterior distribution:
\begin{align*}
	&[\boldsymbol{\zeta}|\boldsymbol{y},\ldots ]\sim N\bigg( \bigg(\frac{\bm{\eta_{\zeta}}^2}{\boldsymbol{\nu^2} + \bm{\eta_{\zeta}}^2}\bigg) \bigg(\boldsymbol{Y}-\bm{T} - \sum_{i=1}^{P}\bm{S}_{i}\bigg),\bigg(\frac{\sigma^2_{y}\boldsymbol{\nu}^2\bm{\eta_{\zeta}}^2}{\boldsymbol{\nu^2} + \bm{\eta_{\zeta}}^2}\bigg)\bigg),\\
	&[\eta^2_{\zeta,t}|\boldsymbol{y},\ldots ] \sim IG\bigg(1,\frac{1}{\psi_{\eta_{\zeta},t}} + \frac{1}{2}\bigg(\frac{\zeta_{t}}{\sigma_{y}}\bigg)^2 \bigg)\\
	&[\psi_{\eta_{\zeta},t}|\boldsymbol{y},\ldots ] \sim IG\bigg(1,\frac{1}{\eta_{\zeta,t}} + \frac{1}{\tau^2_{\zeta}\xi^2_{\zeta,t}} \bigg)\\
	&[\tau^2_{\zeta}|\bm{y},\ldots ] \sim IG\bigg(\frac{N+1}{2}, \sum_{t=1}^{N}\bigg(\frac{1}{\xi^2_{\zeta,t} \psi_{\eta_{\zeta},t}}\bigg) + \frac{1}{\psi_{\tau_{\zeta}}}\bigg) \\
	&[\psi_{\tau_{\zeta}}|\bm{y}] \sim IG\bigg(1, 1 + \frac{1}{\tau^2_{\zeta}}\bigg) \\
	&[\xi^2_{\zeta,t}|\bm{y},\ldots] \sim IG(\bigg(1,\frac{1}{\tau^2_{\zeta}\psi_{\eta_{\zeta},t}} + \frac{1}{\psi_{\xi_{\zeta},t}} \bigg)\\
	&[\psi_{\xi_{\zeta},t}|\bm{y}] \sim IG\bigg(1, 1 + \frac{1}{\xi^2_{\zeta,t}}\bigg) 
\end{align*}

\subsection{Remainders}
Remainder term also decomposes into two components, the time-invariant $\sigma_{y}$, and the time-varying term $\nu_{t}$. For the time-invariant term, we have

\begin{align*}
	\sigma_{y}^2 \sim IG\bigg(\frac{3N + P(N-1) }{2},\frac{1}{2}\bigg(&\bigg(\frac{\bm{Y}-\bm{T} - \sum_{i=1}^{P}\bm{S_{i}} - \zeta}{\bm{\nu}}\bigg)^2 +  \\
	& \sum_{t=1}^{2} \bigg(\frac{T_{t}}{\eta_{T,t}}\bigg)^2
	+ \sum_{t=3}^{N} \bigg(\frac{\Delta^2 T_{t}}{\tau_{T}\eta_{T,t}}\bigg)^2 + \\
	& \sum_{i=1}^{P}\bigg(\bigg(\frac{S_{i,2}}{\eta_{{S_i},2}}\bigg)^2 +  \sum_{t=3}^{k} \bigg(\frac{\Delta^2 S_{i,t}}{\tau_{S_{i}}\eta_{S_{i,t}}} \bigg)^2 + \sum_{k+1}^{N} \bigg(\frac{(1-B)^{k_{i}}S_{i,t}}{\tau_{S_{i}}\eta_{S_{i,t}}} \bigg)^2 \bigg) + \\
	& \sum_{t=1}^{N}\bigg(\frac{\zeta_{t}}{\eta_{\zeta,t}}\bigg)^2\bigg) \bigg)
\end{align*}
For the time-varying term $\bm{\nu}$, we closely follow the sampler proposed by \cite{Kastner_2014}. Define
\begin{align*}
	&\bm{Y}^* = log\bigg(\frac{\bm{Y}-\bm{T} - \sum_{i=1}^{P}\bm{S_{i}} - \bm{\zeta}}{\sigma_{y}}\bigg)^2  \\
	&\bm{h} := log(\bm{\nu}^2).
\end{align*}
By transforming the likelihood, we have a linear system with a non-Gaussian error term, which may be approximated by a Gaussian mixture distribution \cite{omorietal}:
\begin{align*}
    &y_{t}^* = h_{t} + \log(u_{t}^2), & u_t \stackrel{iid}{\sim} N(0,1), \\
    &y_{t}^* \approx h_{t} + \mu_{j_{t}} + \sigma_{j_{t}} u_t, & u_t \stackrel{iid}{\sim} N(0,1), \;\; j_{t} \stackrel{iid}{\sim} \text{Categorical}(\pi) \\
    &h_{t} = \mu + \phi(h_{t-1} - \mu) + \sigma_{\nu}\epsilon & [\epsilon_{t}]\stackrel{iid}{\sim}N(0,1)
\end{align*}
With the approximation of the likelihood, the model becomes conditionally Gaussian, allowing efficient sampling of associated parameter $\boldsymbol{h}, \mu, \phi, \sigma_{\nu}$, and the additional parameter $\bm{j}$.

\section{Simulation Schemes}\label{App:simscheme}
\subsection{Data Generating Scheme 1}
\noindent \textbf{Observation Equation}
\begin{align*}
	&Y_t = T_t + S^{12}_t + S^{40}_t + R_t.
\end{align*}
\textbf{Trend}
\begin{align*}
    &T_t =
    \begin{cases} 
        m_1(0.04)(t - b_1) + c_1, & \text{if } 0 \leq t < b_1, \\
        m_2(0.04)(t - b_2) + c_2, & \text{if } b_1 \leq t < b_1 + b_2, \\
        m_3(0.04)(t - b_3) + c_3, & \text{if } b_1 + b_2 \leq t < b_1 + b_2 + b_3, \\
        m_4(0.04)(t - b_4) + c_4, & \text{if } b_1 + b_2 + b_3 \leq t \leq 500,
    \end{cases}\\
    &[m_1, \ldots, m_4] \stackrel{iid}{\sim} U(-20, 20), \\
	&[c_1, \ldots, c_4] \stackrel{iid}{\sim} U(-10, 10), \\
    &[b_1, \ldots, b_3] \stackrel{iid}{\sim} U(30, 125).
\end{align*}
\bf{Seasonal}
\begin{align*}
    &S^{12}_t = \gamma^{12}_1 \sin\left(\frac{2\pi t}{12}\right) + \gamma^{12}_2\cos\left(\frac{2\pi t}{12}\right)\\  
    &[\gamma^{12}_1,\gamma^{12}_2] \stackrel{iid}{\sim} N(0,4^2),\\
    &S^{40}_t = \gamma^{40}_1 \sin\left(\frac{2\pi t}{40}\right) + \gamma^{40}_2 \cos\left(\frac{2\pi t}{40}\right) \\
    &[\gamma^{40}_1,\gamma^{40}_2] \stackrel{iid}{\sim} N(0,5^2).
\end{align*}
\bf{Error}
\begin{align*}
	[R_{t}] \stackrel{iid}{\sim} N(0,2^2)
\end{align*}
\subsection{Data Generating Scheme 2}
\noindent \textbf{Observation Equation}
\begin{align*}
	&Y_t = T_t + S^{40}_t + R_t.
\end{align*}
\textbf{Trend}
\begin{align*}
	T_{t} &= \frac{m t}{500} \\
	[m] &\sim N(0, 30^2).
\end{align*}
\bf{Seasonal}
\begin{align*}
    S_t &= 
\begin{cases} 
\tilde{v}_1, & \text{if } t \in \{t \mid t = 40k + j, 0 \leq k \leq 8, 1 \leq j \leq 10\}, \\
\tilde{v}_2, & \text{if } t \in \{t \mid t = 40k + j, 0 \leq k \leq 8, 11 \leq j \leq 20\}, \\
\tilde{v}_3, & \text{if } t \in \{t \mid t = 40k + j, 0 \leq k \leq 8, 21 \leq j \leq 30\}, \\
\tilde{v}_4, & \text{if } t \in \{t \mid t = 40k + j, 0 \leq k \leq 8, 31 \leq j \leq 40\},
\end{cases}\\
     &[v_1,\ldots,v_{4}] \stackrel{iid}{\sim} U\left(-8, 8\right), \\
     &\tilde{v}_i = v_i - \frac{1}{4} \sum_{j=1}^4 v_j.
\end{align*}
\bf{Error}
\begin{align*}
	[R_{t}] \stackrel{iid}{\sim} N\bigg(0,\frac{m^2}{100}\bigg).
\end{align*}
\subsection{Data Generating Scheme 3}
\noindent \textbf{Observation Equation}
\begin{align*}
	&Y_t = T_t + S^{50}_t + R_t.
\end{align*}
\textbf{Trend}
\begin{align*}
	&T_{t} = b_0 + b_{1}\frac{t}{500} + b_{2}\bigg(\frac{t}{500}\bigg)^{2} + b_{3}\bigg(\frac{t}{500}\bigg)^{3}\\ 
	&[b_0] \sim U(-15,15), \\ 
	&[b_1,\ldots,b_{4}] \stackrel{iid}{\sim} N(0,20^2).
\end{align*}
\textbf{Seasonal}
\begin{align*}
    & S^{50}_t = \gamma^{50}_1 \sin\left(\frac{2\pi t}{50}\right) + \gamma^{50}_2 \cos\left(\frac{2\pi t}{50}\right)\\ 
    & [\gamma^{50}_1,\gamma^{50}_2] \stackrel{iid}{\sim} N(0,5^2).
\end{align*}
\bf{Error}
\begin{align*}
	&[R_{t}] \stackrel{iid}{\sim} N(0,exp(h_t)) \\
	&h_{t} = 2.5 + 0.98(h_{t-1} - 2.5) + 0.2 \epsilon_t \\
	&[\epsilon_{t}] \stackrel{iid}{\sim} N(0,1)
\end{align*}
\subsection{Data Generating Scheme 4}
\noindent \textbf{Observation Equation}
\begin{align*}
	&Y_t = T_t + S^{12}_t + S^{50}_t + O_{t} + R_t, & t = 1, \ldots, 500.
\end{align*}
\textbf{Trend}
\begin{align*}
    T_t &=
    \begin{cases} 
        m_1(0.04)(t - b_1) + c_1, & \text{if } 0 \leq t < b_1, \\
        m_2(0.04)(t - b_2) + c_2, & \text{if } b_1 \leq t < b_1 + b_2, \\
        m_3(0.04)(t - b_3) + c_3, & \text{if } b_1 + b_2 \leq t < b_1 + b_2 + b_3, \\
        m_4(0.04)(t - b_4) + c_4, & \text{if } b_1 + b_2 + b_3 \leq t \leq 500,
    \end{cases}\\
    &[m_1, \ldots, m_4] \stackrel{iid}{\sim} U(-20, 20), \\
	&[c_1, \ldots, c_4] \stackrel{iid}{\sim} U(-10, 10), \\
    &[b_1, \ldots, b_3] \stackrel{iid}{\sim} U(30, 125).
\end{align*}
\bf{Seasonal}
\begin{align*}
    &S^{12}_t = \gamma^{12}_1 \sin\left(\frac{2\pi t}{12}\right) + \gamma^{12}_2\cos\left(\frac{2\pi t}{12}\right)\\
    & [\gamma^{12}_1,\gamma^{12}_2] \stackrel{iid}{\sim} N(0,4^2),\\
    & S^{50}_t = 
\begin{cases} 
\tilde{v}_1, & \text{if } t \in \{t \mid t = 40k + j, 0 \leq k \leq 8, 1 \leq j \leq 10\}, \\
\tilde{v}_2, & \text{if } t \in \{t \mid t = 40k + j, 0 \leq k \leq 8, 11 \leq j \leq 20\}, \\
\tilde{v}_3, & \text{if } t \in \{t \mid t = 40k + j, 0 \leq k \leq 8, 21 \leq j \leq 30\}, \\
\tilde{v}_4, & \text{if } t \in \{t \mid t = 40k + j, 0 \leq k \leq 8, 31 \leq j \leq 40\},
\end{cases}\\
     &[v_1,\ldots,v_{4}] \stackrel{iid}{\sim} U\left(-20, 20\right), \\
     &\tilde{v}_i = v_i - \frac{1}{4} \sum_{j=1}^4 v_j
\end{align*}
\bf{Outlier}
\begin{align*}
& O_t = 
	\begin{cases} 
		a_t, & \text{if } t \in \{j_1, j_2, \ldots, j_{l}\}, \\
		0, & \text{if } t \not\in \{j_1, j_2, \ldots, j_{l}\}, \\
	\end{cases} \\
& a_t \stackrel{iid}{\sim} N(0, 15^2), \\
& l \sim Pois(5), \\
& [j_1, j_2, \ldots, j_l] \sim U(1, 500) \text{ without replacement}.
\end{align*}
\bf{Error}
\begin{align*}
	&[R_{t}] \stackrel{iid}{\sim} N(0,exp(h_t)) \\
	&h_{t} = 0.8 + 0.98(h_{t-1} - 2.5) + 0.2 \epsilon_t \\
	&[\epsilon_{t}] \stackrel{iid}{\sim} N(0,1)
\end{align*}
\section{Additional Figures}\label{App:addfigs}
\begin{figure}[!ht]
    \centering 
    \subfigure[DGP 1]{%
        \includegraphics[width=0.45\linewidth]{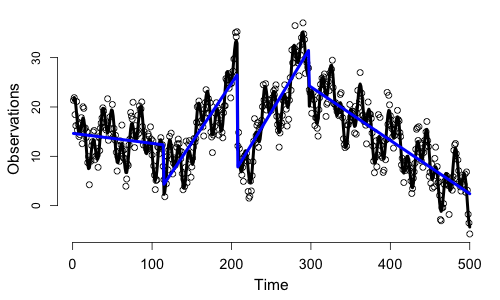}
        \label{fig:DGP1}
    }
    \subfigure[DGP 2]{%
        \includegraphics[width=0.45\linewidth]{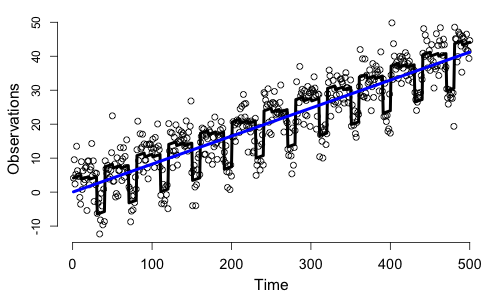}
        \label{fig:DGP2}
    }
    \subfigure[DGP 3]{%
        \includegraphics[width=0.45\linewidth]{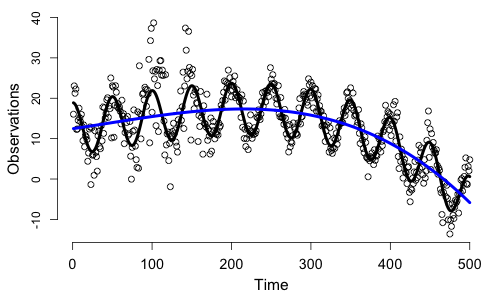}
        \label{fig:DGP3}
    }\hfill
    \subfigure[DGP 4]{%
        \includegraphics[width=0.45\linewidth]{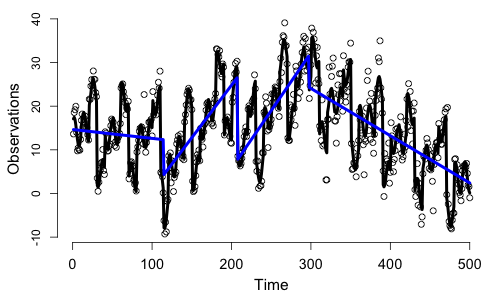}
        \label{fig:DGP4}
    }
    \caption{Synthetic time series generated by adding a trend component, $T_{t}$, seasonal components, $S_{t}$, and remainders $R_{t}$ based on the descriptions in Table~\ref{tab:simulation_descriptions}. $T_{t}$ are drawn in blue and $T_{t} + S_{t}$ are drawn in black. Figures represent one replication of each DGP.}
    \label{fig:simulation_example}
\end{figure}
\begin{figure}[!ht]
    \centering 
    \subfigure[Observed Data]{%
        \includegraphics[width=0.4\linewidth]{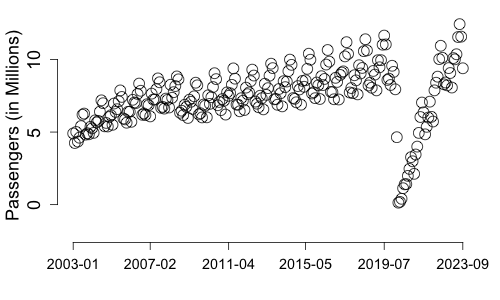}
        \label{fig:airline_y}
    }
    \subfigure[Trend Component]{%
        \includegraphics[width=0.4\linewidth]{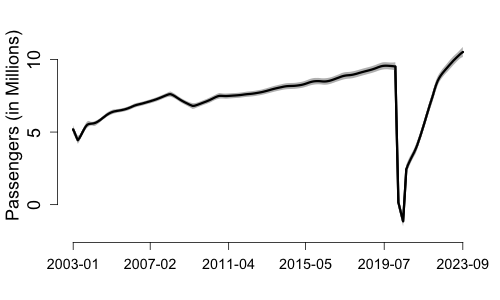}
        \label{fig:airline_t}
    }
    \subfigure[Seasonality Component]{%
        \includegraphics[width=0.4\linewidth]{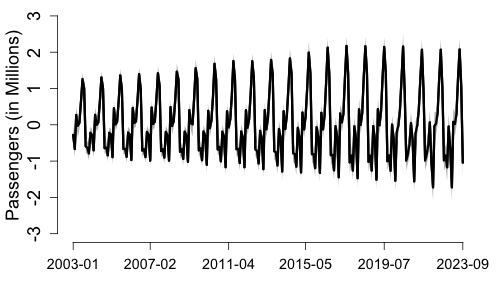}
        \label{fig:airline_s12}
    }
    \subfigure[Remainder]{%
        \includegraphics[width=0.4\linewidth]{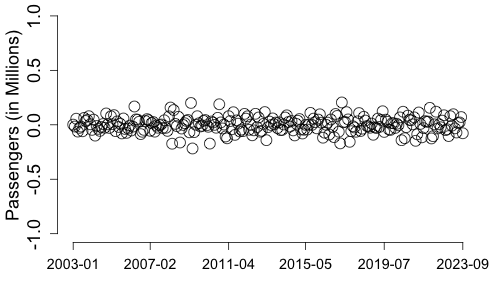}
        \label{fig:airline_rem}
    }
    \caption{Monthly international airline traffic in the U.S from 2003 to 2023 \ref{fig:airline_y}, and its Trend \ref{fig:airline_t}, Seasonality \ref{fig:airline_s12}, and remainder Figure~\ref{fig:airline_rem} decomposition based on the proposed model, BASTION. 95\% credible regions are generated and shaded in grey for the trend and seasonality components.}
    \label{fig:airline_BASTION}
\end{figure}
\begin{figure}[!ht]
    \centering 
    \subfigure[Trend Estimate and observed data]{%
        \includegraphics[width=0.4\linewidth]{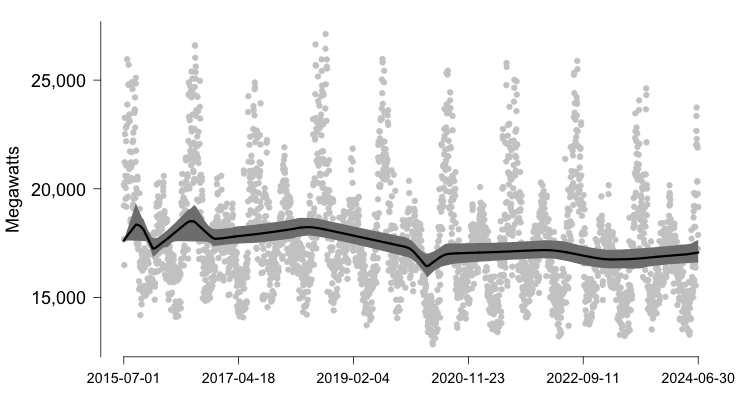}
        \label{fig:elec_trend}
    }
    \subfigure[Volatility Estimate]{%
        \includegraphics[width=0.4\linewidth]{Figures/Electricity/BASTION_vol.PNG}
        \label{fig:elec_vol2}
    }
    \subfigure[Weekly Seasonality]{%
        \includegraphics[width=0.4\linewidth]{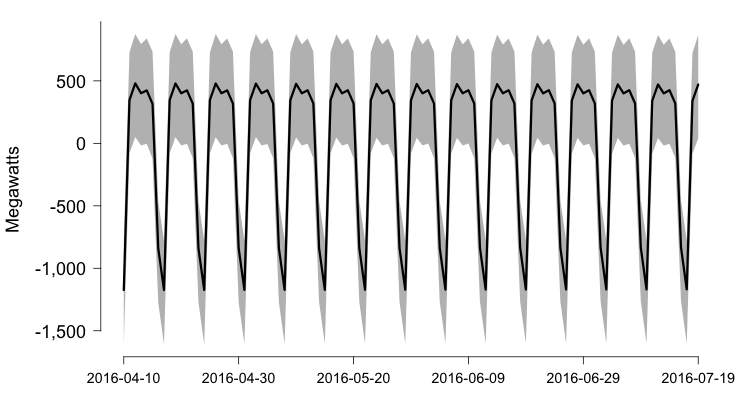}
        \label{fig:elec_week}
    }
    \subfigure[Yearly Seasonality]{%
        \includegraphics[width=0.45\linewidth]{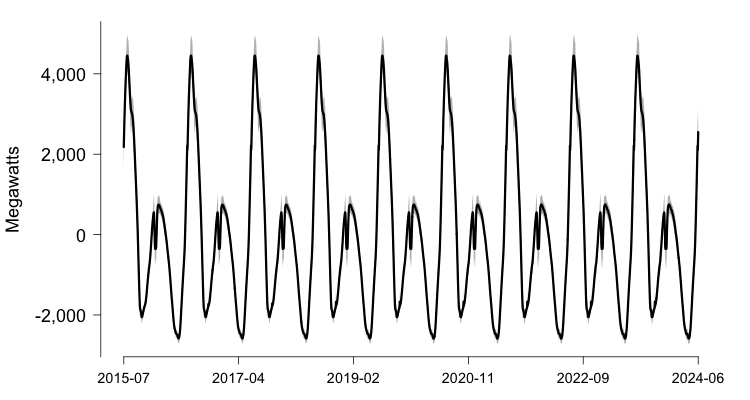}
        \label{fig:elec_year}
    }
    \caption{Decomposition of the trend \ref{fig:elec_trend}, volatility \ref{fig:elec_vol}, weekly seasonality \ref{fig:elec_week}, and yearly seasonality \ref{fig:elec_year} based on BASTION for daily average electricity demand from 2015-07-01 to 2024-06-30. 95\% credible regions are drawn in dark grey.}
    \label{fig:elec_decom}
\end{figure}
\end{document}